\documentclass[twocolumn]{aastex701}
\usepackage[utf8]{inputenc}

\definecolor{newred}{rgb}{0.1,0.6,0.75}
\definecolor{grey}{rgb}{0.2,0.27,0.57}
\definecolor{freeblue}{rgb}{0.2,0.25,0.45}
\definecolor{freeblue2}{rgb}{0.2,0.25,0.4}
\definecolor{blue}{rgb}{0.45,0.41,0.98}
\definecolor{myblue}{rgb}{0.1,0.5,0.5}
\definecolor{guillaume}{rgb}{0.,0.5,0.65}
\definecolor{louloublue}{rgb}{0.2,0.2,0.65}
\definecolor{loulougreen}{rgb}{0.1,0.35,0.75}
\definecolor{violet}{rgb}{0.5,0.,0.5}

\newcolumntype{i}{>{\scriptsize}r}

\shortauthors{Molina et al.}

\graphicspath{{./}{figs/}}

\begin{document}

\shorttitle{Periodicities in the 46-yr Radio Light Curves of 83 Blazars}
\title{A Search for Supermassive Black Hole Binary Candidates in 46-Year Radio Light Curves of 83 Blazars}
\color{black}
\correspondingauthor{Anthony Readhead}
\email{acr@caltech.edu}
\color{black}

\author[0009-0000-9963-6874]{B. Molina}
\email{brian.leftraru@gmail.com}   
\affiliation{CePIA, Astronomy Department, Universidad de Concepci\'on,  Casilla 160-C, Concepci\'on, Chile}

\author[0000-0001-7016-1692]{P. Mr{\'o}z}
\email{Pmroz@astrouw.edu.pl}   
\affiliation{Astronomical Observatory, University of Warsaw, Al. Ujazdowskie 4, 00-478 Warszawa, Poland}

\author[0000-0001-5957-1412]{P. V.~De la Parra}
\email{phvergara@udec.cl}   
\affiliation{CePIA, Astronomy Department, Universidad de Concepci\'on,  Casilla 160-C, Concepci\'on, Chile}

\author[0000-0001-9152-961X]{A.~C.~S.~Readhead}
\email{acr@caltech.edu}
\affiliation{Owens Valley Radio Observatory, California Institute of Technology,  Pasadena, CA 91125, USA}
\affiliation{Institute of Astrophysics, Foundation for Research and Technology-Hellas, GR-71110 Heraklion, Greece}

\author[0000-0002-6369-6266]{T. Surti}
\email{tsurti@caltech.edu}  
\affiliation{Owens Valley Radio Observatory, California Institute of Technology,  Pasadena, CA 91125, USA}

\author[0000-0003-2483-2103]{M. F. Aller}
\email{mfa@umich.edu}   
\affiliation{Department of Astronomy, University of Michigan, 323 West Hall, 1085 S. University Avenue, Ann Arbor, MI 48109, USA}

\author[0000-0001-5623-0065]{J. D. Scargle}
\email{jeffscargle@gmail.com}
\affiliation{Astrobiology and Space Science Division, NASA Ames Research Center (retired)}

\author[0000-0001-5704-271X]{R. A. Reeves}
\email{rreevesd@gmail.com}  
\affiliation{CePIA, Astronomy Department, Universidad de Concepci\'on,  Casilla 160-C, Concepci\'on, Chile}

\author[0000-0003-1945-1840]{H. Aller}
\email{haller@umich.edu}   
\affiliation{Department of Astronomy, University of Michigan, 323 West Hall, 1085 S. University Avenue, Ann Arbor, MI 48109, USA}

\author[0000-0003-0936-8488]{M. C.  Begelman}
\email{mitch@jila.colorado.edu}
\affiliation{JILA, University of Colorado and National Institute of Standards and Technology, 440 UCB, Boulder, CO 80309-0440, USA} 

\author[0000-0002-1854-5506]{R. D. Blandford}
\email{rdb3@stanford.edu}   
\affiliation{Kavli Institute for Particle Astrophysics and Cosmology, Department of Physics,
Stanford University, Stanford, CA 94305, USA}

\author[0000-0002-5770-2666]{Y. Ding}
\email{yding@caltech.edu}   
\affiliation{Cahill Center for Astronomy and Astrophysics, California Institute of Technology, Pasadena, CA 91125, USA}

\author[0000-0002-3168-0139]{M. J. Graham}
\email{mjg@caltech.edu}   
\affiliation{Division of Physics, Mathematics, and Astronomy, California Institute of Technology, Pasadena, CA 91125, USA}

\author[0000-0002-4226-8959]{F. Harrison}
\email{fiona@srl.caltech.edu}   
\affiliation{Cahill Center for Astronomy and Astrophysics, California Institute of Technology, Pasadena, CA 91125, USA}

\author[0000-0002-2024-8199]{T. Hovatta}
\email{talvikki.hovatta@utu.fi}   
\affiliation{Finnish Centre for Astronomy with ESO (FINCA), University of Turku, FI-20014 University of Turku, Finland}
\affiliation{Aalto University Department of Electronics and Nanoengineering, PL~15500, FI-00076 Espoo, Finland}
\affiliation{Aalto University Mets\"ahovi Radio Observatory,  Mets\"ahovintie 114, 02540 Kylm\"al\"a, Finland}

\author[0000-0001-9200-4006]{I. Liodakis}
\email{yannis.liodakis@gmail.com}   
\affiliation{Institute of Astrophysics, Foundation for Research and Technology-Hellas, GR-71110 Heraklion, Greece}

\author[0000-0003-1315-3412]{M. L. Lister}
\email{mlister@purdue.edu}   
\affiliation{Department of Physics and Astronomy, Purdue University, 525 Northwestern Avenue, West Lafayette, IN 47907, USA}

\author[0000-0002-5491-5244]{W. Max-Moerbeck} 
\email{wmax@das.uchile.cl}   
\affiliation{Departamento de Astronomía, Universidad de Chile, Camino El Observatorio 1515, Las Condes, Santiago, Chile}

\author[0000-0002-0870-1368]{V. Pavlidou} 
\email{pavlidou@physics.uoc.gr}  
\affiliation{Department of Physics and Institute of Theoretical and Computational Physics, University of Crete, 71003 Heraklion, Greece}
\affiliation{Institute of Astrophysics, Foundation for Research and Technology-Hellas, GR-71110 Heraklion, Greece}

\author[0000-0001-5213-6231]{T. J. Pearson}
\email{tjp@astro.caltech.edu}   
\affiliation{Owens Valley Radio Observatory, California Institute of Technology,  Pasadena, CA 91125, USA}

\author[0000-0002-7252-5485]{V. Ravi}
\email{vikram@astro.caltech.edu}   
\affiliation{Owens Valley Radio Observatory, California Institute of Technology,  Pasadena, CA 91125, USA}

\author[0000-0002-9545-7286]{A. G. Sullivan}
\email{andrew.sullivan@stanford.edu}   
\affiliation{Kavli Institute for Particle Astrophysics and Cosmology, Department of Physics,
Stanford University, Stanford, CA 94305, USA}

\author[0009-0004-2614-830X]{A. Synani} 
\email{akyvsyn@physics.uoc.gr}   
\affiliation{Department of Physics and Institute of Theoretical and Computational Physics, University of Crete, 71003 Heraklion, Greece}
\affiliation{Institute of Astrophysics, Foundation for Research and Technology-Hellas, GR-71110 Heraklion, Greece}

\author[0000-0002-8831-2038]{K. Tassis}
\email{tassis@physics.uoc.gr}   
\affiliation{Department of Physics and Institute of Theoretical and Computational Physics, University of Crete, 71003 Heraklion, Greece}
\affiliation{Institute of Astrophysics, Foundation for Research and Technology-Hellas, GR-71110 Heraklion, Greece}

\author[0000-0001-7662-2576]{S. E. Tremblay}
\email{strembla@nrao.edu}   
\affiliation{National Radio Astronomy Observatory, 1011 Lopez Road, Socorro, NM 87801, USA}

\author[0000-0001-7470-3321]{J. A. Zensus}
\email{azensus@mpifr-bonn.mpg.de}   
\affiliation{Max-Planck-Institut f\"ur Radioastronomie, Auf dem H\"ugel 69, D-53121 Bonn, Germany}

\begin{abstract}
The combined University of Michigan Radio Astronomy Observatory (UMRAO) and Owens Valley Radio Observatory (OVRO) blazar monitoring programs at 14.5/15 GHz provide uninterrupted light curves of $\sim~46-50$ yr duration for 83 blazars, selected from amongst the brightest and most rapidly flaring blazars north of declination $-20^\circ$.  In a search for supermassive black hole binary (SMBHB) candidates, we carried out tests for periodic variability using generalized Lomb-Scargle (GLS), weighted wavelet-Z (WWZ), and sine-wave fitting (SWF) analyses of this sample.  We used simulations to test the effects of the power law spectrum of the power spectral density (PSD) on our findings, and show that the irregular sampling in the observed light curves has very little effect on the GLS spectra.   Apparent periodicities and putative harmonics appear in all 83 of the GLS spectra of the blazars in our sample.  We tested the reality of these apparent periodicities and harmonics with simulations, and found that in the overwhelming majority of cases they are due to the steep slope of the PSD, and should therefore be treated with great caution.  We find one  new SMBHB candidate: PKS 1309+1154, which exhibits a 17.9 year periodicity. The fraction of SMBHB candidates in our sample is $2.4_{-0.8}^{+3.2}\%$.
\end{abstract}


\section{Introduction}
\label{sec:intro}

Periodicities in the light curves of objects powered by black holes in both stellar mass systems and active galactic nuclei (AGN)  are of great scientific interest.  They provide a direct probe into the structure of these systems on scales that cannot be probed by imaging with even the highest resolution available to astronomy, i.e., very long baseline interferometry (VLBI), which provides resolution as fine as 20 microarcseconds at sub-millimeter wavelengths \citep{2019ApJ...875L...4E}.  Unfortunately, such light curves are much more difficult to interpret than  images of resolved structures.

Periodicities in AGN could well be indicative of supermassive black hole binaries (SMBHBs), which are thought to be responsible for  a stochastic background of gravitational waves (GW) with periods of months to years \citep{2023ApJ...951L...8A,epta:2023,2023MNRAS.519.3976M,2025arXiv250816534A}. 

Most blazar light curves are dominated by large flares, but recently, in the Owens Valley Radio Observatory (OVRO) 40 m Telescope monitoring program \citep{2011ApJS..194...29R}, two blazars have been found in which sinusoidal variations of constant period dominate their light curves much of the time: PKS J0805--0111 and PKS 2131--021   
(\citealt{2021MNRAS.506.3791R};
\citealt{2021RAA....21...75R};
\citealt{2022ApJ...926L..35O} [Paper 1];
\citealt{2025ApJ...985...59K} [Paper 2];
\citealt{2025ApJ...987..191D} [Paper 3];
\citealt{2025arXiv250404278H} [Paper 4]).
Papers 1--4 carried out tests using simulations that take into account the steepness of the power spectral density (PSD) and the probability density function (PDF), and showed that the sinusoidal variations are unlikely to be due to the random nature of blazar flares.  PKS J0805--0111 and PKS 2131--021  are, therefore, strong candidates for supermassive black hole binaries (SMBHBs).

The University of Michigan Radio Astronomical Observatory (UMRAO) 26 m Telescope was dedicated to a blazar monitoring program from 1979 to 2013 \citep{1985ApJS...59..513A,2014ApJ...791...53A}, at the primary frequency of 14.5 GHz. The OVRO 40 m Telescope has been dedicated to an AGN monitoring program of \hbox{$\sim1830$} blazars at 15 GHz since 2008.  The 83 blazars that are common to these two programs and that were observed continuously from 1979 to 2025 are the subject of this paper.

In this paper we report the discovery of strong sinusoidal emission in PKS 1309+1154,\footnote{PKS J1309+1154 is a BL Lac object that has no reliable redshift value (see \S\ref{sec:J1309optical}).} with a period of 17.9 years, based on combined observations from the UMRAO and the OVRO that span the epoch range 1975-2024. Because this window covers only 2.5 periods, PKS 1309+1154 is  not as strong an SMBHB  candidate as PKS J0805--0111 or PKS 2131--021.

Periodicities and possible harmonics in blazar light curves have been discussed in a number of papers \citep[e.g.,][]{2000ApJ...545..758R,2006ApJ...650..749L,2013MNRAS.434.3487A,2014MNRAS.443...58W,2015ARep...59..851B,2017ApJ...847....7B,2017ChA&A..41...42W,2021MNRAS.501.5997T,2022RAA....22k5001D,2023RAA....23i5010L,2023MNRAS.523L..52B,2023arXiv231212623S,2024MNRAS.531.3927M,2025A&A...698A.265L}.  We carried out tests for periodic variability of sources in the 83 blazars using generalized Lomb-Scargle (GLS), weighted wavelet-Z (WWZ), and sine-wave fitting (SWF) analyses. Our GLS spectrum analysis \citep{1976Ap&SS..39..447L,1982ApJ...263..835S,2009AandA...496..577Z} reveals similar apparent periodicities and their harmonics in the light curves of \textit{all} 83 blazars in our sample.  

In this paper we show that, in the overwhelming majority of cases, these periodicities and apparent harmonics are  likely due to the random nature of blazar flares caused by the steep PSD spectrum, and that the observing cadence does not matter provided the objects are well-sampled on the timescale of the periodicity.

\begin{figure}[t]
   \centering
   \includegraphics[width=\linewidth]{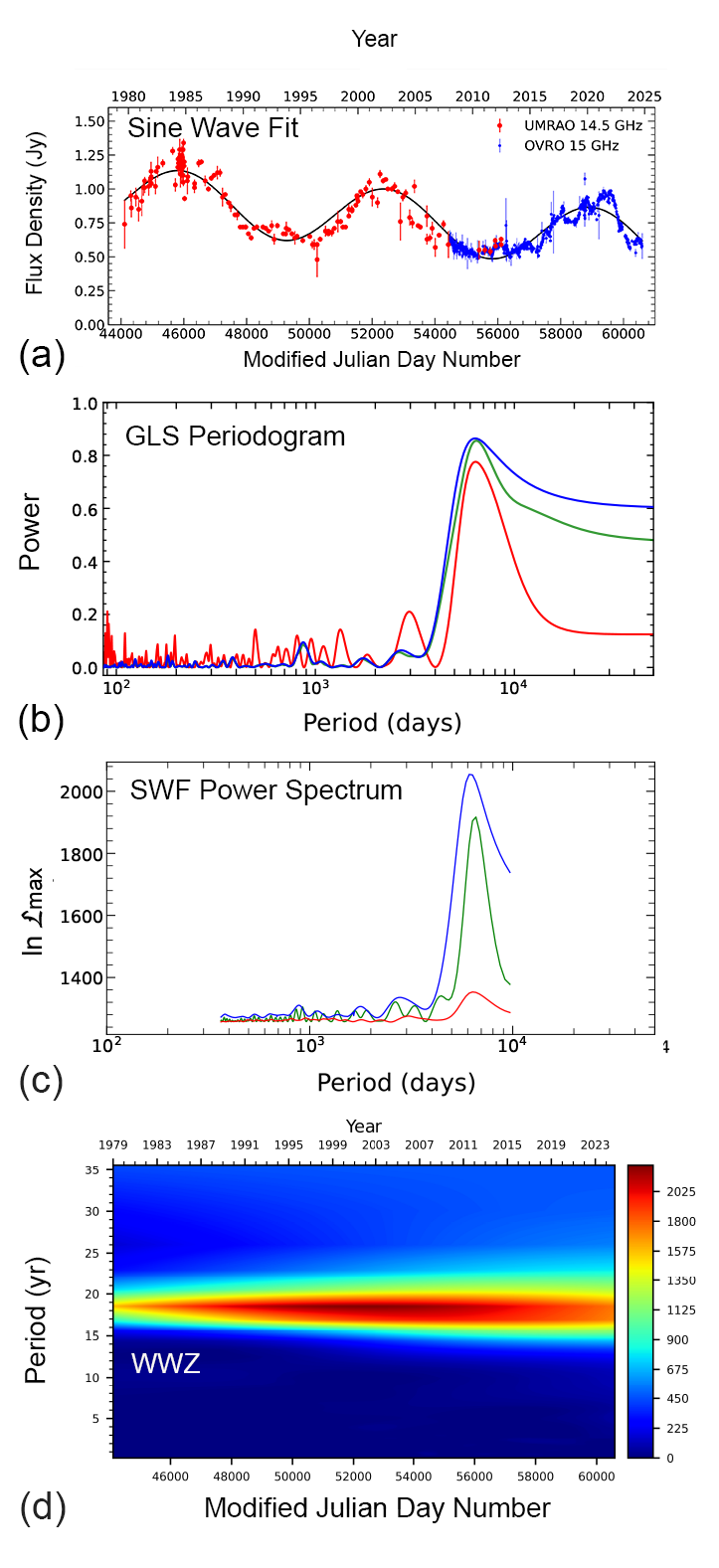}
   \caption{PKS J1309+1154: (a) The combined UMRAO and OVRO light curve.  Red and blue symbols denote the UMRAO and OVRO data, respectively.  The fitted black sine wave includes a linear downward trend and has period of 6551 days.   (b) The GLS periodogram: red, blue, and green curves are the UMRAO, OVRO, and combined UMRAO+OVRO spectra, respectively. (c) The SWF power spectrum. (d) The WWZ wavelet analysis showing only one prominent feature over 46 years: color bar: WWZ power $Z$ (unitless, linear).}
   \label{plt:lightcurve1}
\end{figure}

\newcommand{\colnumoffset}{0.28em}
\newcommand{\colnummini}[1]{%
  \rlap{\hspace{\colnumoffset}\textsuperscript{\smash{\raisebox{0.15ex}{\scalebox{0.78}{(\,#1\,)}}}}}
}

\begin{deluxetable*}{lcccc@{\hskip 8mm}lcccc}
\tablecaption{The Combined UMRAO+OVRO Sample of 83 BLazars\label{tab:sample}}
\tablehead{
\colhead{\shortstack[c]{J2000\colnummini{1}\\Name}} &
\colhead{\shortstack[c]{B1950/\colnummini{2}\\C.\ Name}} &
\colhead{\shortstack[c]{RA\colnummini{3}\\(deg)}} &
\colhead{\shortstack[c]{Dec\colnummini{4}\\(deg)}} &
\colhead{\shortstack[c]{$z$\colnummini{5}\\\vphantom{X}}} &
\colhead{\shortstack[c]{J2000\colnummini{6}\\Name}} &
\colhead{\shortstack[c]{B1950/\colnummini{7}\\C.\ Name}} &
\colhead{\shortstack[c]{RA\colnummini{8}\\(deg)}} &
\colhead{\shortstack[c]{Dec\colnummini{9}\\(deg)}} &
\colhead{\shortstack[c]{$z$\colnummini{10}\\\vphantom{X}}}
}
\startdata
J0010+1058 & IIIZW2 & 2.629 & 10.975 & 0.089 & J1217+3007 & 1215+303 & 184.467 & 30.117 & 0.130 \\
J0019+7327 & 0016+731 & 4.941 & 73.458 & 1.781 & J1221+2813 & 1219+285 & 185.382 & 28.233 & 0.102 \\
J0050-0929 & 0048-097 & 12.672 & -9.485 & 0.635 & J1224+2122 & 1222+216 & 186.227 & 21.380 & 0.433 \\
J0102+5824 & 0059+581 & 15.691 & 58.403 & 0.644 & J1229+0203 & 3C273 & 187.278 & 2.052 & 0.158 \\
J0108+0135 & 0106+013 & 17.162 & 1.583 & 2.109 & J1256-0547 & 3C279 & 194.047 & -5.789 & 0.536 \\
J0111+3906 & 0108+388 & 17.905 & 39.108 & 0.668 & J1305-1033 & 1302-102 & 196.388 & -10.555 & 0.278 \\
J0112+2244 & 0109+224 & 18.024 & 22.744 & 0.265 & J1309+1154 & 1307+121 & 197.391 & 11.907 & U \\
J0136+4751 & 0133+476 & 24.244 & 47.858 & 0.859 & J1310+3220 & 1308+326 & 197.619 & 32.345 & 0.995 \\
J0204+1514 & 0202+149 & 31.210 & 15.236 & 0.405 & J1337-1257 & 1334-127 & 204.416 & -12.957 & 0.539 \\
J0217+7349 & 0212+735 & 34.378 & 73.826 & 2.346 & J1415+1320 & 1413+135 & 213.995 & 13.340 & 0.247 \\
J0217+0144 & 0215+015 & 34.454 & 1.747 & 1.715 & J1419+5423 & 1418+546 & 214.944 & 54.387 & 0.152 \\
J0228+6721& 0224+671 & 37.209 & 67.351 & 0.523 & J1512-0905 & 1510-089 & 228.211 & -9.100 & 0.360 \\
J0237+2848 & 0234+285 & 39.468 & 28.802 & 1.206 & J1540+1447 & 1538+149 & 235.206 & 14.796 & 0.605 \\
J0238+1636 & 0235+164 & 39.662 & 16.616 & 0.940 & J1555+1111 & 1553+113 & 238.929 & 11.190 & 0.49 \\
J0259+0747 & 0256+075 & 44.863 & 7.794 & 0.893 & J1613+3412 & 1611+343 & 243.421 & 34.213 & 1.399 \\
J0303+4716& 0300+471  & 45.897 & 47.271 & U & J1635+3808 & 1633+38 & 248.815 & 38.134 & 1.815 \\
J0309+1029 & 0306+102 & 47.265 & 10.488 & 0.862 & J1642+6856 & 1642+690 & 250.533 & 68.944 & 0.751 \\
J0319+4130 & 3C84 & 49.951 & 41.512 & 0.018 & J1642+3948 & 3C345 & 250.745 & 39.810 & 0.593 \\
J0336+3218& 0333+321 & 54.125 & 32.308 & 1.259 & J1653+3945 & Mrk 501 & 253.468 & 39.760 & 0.034 \\
J0339-0146 & CTA26 & 54.879 & -1.776 & 0.847 & J1719+1745 & 1717+178 & 259.804 & 17.752 & 0.137 \\
J0418+3801 & 0415+379  & 64.589 & 38.027 & 0.049 & J1733-1304 & NRAO 530 & 263.261 & -13.080 & 0.902 \\
J0423-0120 & 0420-014 & 65.816 & -1.343 & 0.913 & J1740+5211 & 1739+522 & 265.154 & 52.195 & 1.379 \\
J0424+0036 & 0422+004& 66.195 & 0.602 & 0.268 & J1743-0350 & 1741-038 & 265.995 & -3.834 & 1.054 \\
J0433+0521 & 3C120 & 68.296 & 5.354 & 0.033 & J1748+7005 & 1749+701 & 267.137 & 70.097 & 0.770 \\
J0501-0159 & 0458-020 & 75.303 & -1.987 & 2.286 & J1751+0939 & 1749+096 & 267.887 & 9.650 & 0.320 \\
J0530+1331 & 0528+134 & 82.735 & 13.532 & 2.070 & J1800+7828 & 1803+784 & 270.190 & 78.468 & 0.680 \\
J0607-0834 & 0605-085 & 91.999 & -8.581 & 0.872 & J1806+6949 & 3C371 & 271.711 & 69.824 & 0.051 \\
J0609-1542 & 0607-157 & 92.421 & -15.711 & 0.324 & J1824+5651 & 1823+568 & 276.029 & 56.850 & 0.663 \\
J0721+7120 & 0716+714 & 110.473 & 71.343 & U & J1927+7358 & 1928+738 & 291.952 & 73.967 & 0.302 \\
J0730-1141 & 0727-115 & 112.580 & -11.687 & 1.591 & J2005+7752 & 2007+777 & 301.379 & 77.879 & 0.342 \\
J0738+1742 & 0735+178 & 114.531 & 17.705 & U & J2007+4029 & 2005+403 & 301.937 & 40.497 & 1.736 \\
J0739+0137 & 0736+017 & 114.825 & 1.618 & 0.189 & J2022+6136 & 2021+614 & 305.528 & 61.616 & 0.227 \\
J0757+0956 & 0754+100 & 119.278 & 9.943 & 0.266 & J2123+0535 & 2121+053 & 320.935 & 5.589 & 1.941 \\
J0808+4950 & 0804+499 & 122.165 & 49.843 & 1.435 & J2134-0153 & 2131-021 & 323.543 & -1.888 & 1.283 \\
J0818+4222 & 0814+425 & 124.567 & 42.379 & U & J2158-1501 & 2155-152 & 329.526 & -15.019 & 0.672 \\
J0831+0429 & 0829+046 & 127.954 & 4.494 & 0.175 & J2225-0457 & 3C446  & 336.447 & -4.950 & 1.404 \\
J0854+2006 & OJ287 & 133.704 & 20.108 & 0.306 & J2232+1143 & CTA102 & 338.152 & 11.731 & 1.037 \\
J0958+6533 & 0954+658 & 149.697 & 65.565 & 0.367 & J2236+2828 & 2234+282 & 339.094 & 28.483 & 0.795 \\
J1058+0133 & 1055+018 & 164.623 & 1.566 & 0.891 & J2253+1608 & 3C454.3 & 343.491 & 16.148 & 0.859 \\
J1104+3812 & 1101+384 & 166.114 & 38.209 & 0.030 & J2257+0743 & 2254+074 & 344.322 & 7.720 & 0.190 \\
J1150+2417 & 1147+245 & 177.580 & 24.298 & 0.209 & J2348-1631 & 2345-167 & 357.011 & -16.520 & 0.576 \\
J1159+2914 & 1156+295 & 179.883 & 29.245 & 0.725 & \dots & \dots & \dots & \dots & \dots \\
\enddata
\tablecomments{``U'' indicates unknown redshift.}
\end{deluxetable*}

\section{Radio Observations}\label{sec:radio}

The UMRAO blazar monitoring program operated at frequencies 4.8 GHz, 8 GHz, and 14.5 GHz and 
 included measurements of both total flux density and polarization. Only the total flux density measurements at 14.5 GHz, the primary frequency throughout most of the UMRAO program, are included here. These observations were made using a system of dual, rotating, linearly-polarized feed horns symmetrically placed about the parabola’s prime focus which fed a broadband, uncooled high electron mobility transistor (HEMT) amplifier with a bandwidth of 1.68 GHz. An on--on observing technique which alternated beams on the source was employed. Each daily measurement is the average of a series of these individual on--on measurements taken over an interval of 30--40 minutes. The standard deviation associated with each daily flux density observation was computed from the system noise temperature and the number of individual on–on measurements made on the particular day. The standard error estimates include the effects of measurement noise, the errors introduced by uncertainties in the pointing corrections applied to the observations, and the uncertainties in determining the antenna gain as a function of time; these include  both random and systematic contributions. The adopted flux-density scale is based on \citet{Baars1977} and uses Cassiopeia A (3C 461)  as the primary flux density standard. A secondary calibrator, taken from a grid of nearby sources, was observed every 1.5 to 3 hours to verify the stability of the gain and the accuracy of the pointing. At 14.5 GHz typically 24 program sources were observed daily, and highest priority in selecting which sources to include in each observing run (generally 48 hours in duration) was given to measurements of sources exceeding 400 mJy in total flux density in order to provide an adequate signal-to-noise ratio in the polarization measurement. The observing cadence for an individual AGN was determined based on the variability state (flaring or quiescent) at the time of the observation and typically ranged from once a week to once a month for the well-observed AGN included in the sample presented here. As a result, the time windows used in the analysis presented here vary from source to source. The shortest is 18 years, and the median value for the sample is 32 years. Observations were made around-the-clock under automatic telescope control, with occasional data gaps due to poor weather, annual source proximity to the sun, and equipment failures.

The OVRO monitoring program operated on a cadence of once or twice a week.    Occasional gaps exist in the data due to weather conditions or hardware problems. The telescope is equipped with a cryogenic pseudo-correlation receiver.  At the start of the monitoring program, this was a Dicke-switched system with bandpass  from  13.5 GHz  to 16.5 GHz and center frequency 15.0 GHz.  In 2014 this was changed to a correlation receiver with bandpass from 13.2 GHz to  17.8 GHz and center frequency 15.5 GHz. The receiver has dual-beam switching to remove atmospheric contributions and background emission, as described by \citet{1989ApJ...346..566R}.  At this radio frequency the observations are confusion-limited, due to the double-switching technique which combines three separate fields,  and the resulting flux density uncertainty is  $\sim$3--4 mJy. The data reduction and calibration processes used to produce the light curves are described in \citet{2011ApJS..194...29R}.

\section{SMBHB Candidates in the Sample of 83 Blazars}\label{sec:qpos}

The 83 blazars in the combined UMRAO+OVRO sample are listed in Table \ref{tab:sample}. In testing for periodicities in the light curves of each of these objects, we used GLS periodograms, and  weighted wavelet Z-tansforms (WWZ) \citep{1996AJ....112.1709F},  to  identify potential  SMBHB candidates showing periodicities.
 In addition,  we have developed the SWF approach, described in Appendix \ref{sec:sine}, to search for potential SMBHB candidates revealed through sinusoidal variations in the light curves.  We used all three of these methods to search for periodicities in each source in the UMRAO, the OVRO, and the UMRAO+OVRO light curves, separately. 

The results from all three methods are consistent: only one source, PKS J1309+1154, is a potential SMBHB candidate. Its light curve is shown in Fig.~\ref{plt:lightcurve1}(a), where the black curve shows the least squares fitted  sine wave combined with a linear trend of $-0.00755$ Jy yr$^{-1}$  to the combined data set. The  period is $P=6551\pm26$ days,  determined as described in Appendix \ref{sec:sine}.  The GLS analysis of the 83 blazars in our sample shows that PKS J1309+1154 (Fig.   \ref{plt:lightcurve1}(b)) is the only case in which the strongest GLS peak  occurs at the same period in the GLS spectrum in the UMRAO, OVRO and UMRAO+OVRO light curves. Likewise, the SWF analysis (Fig.   \ref{plt:lightcurve1}(c) and Table \ref{tab:sine} in Appendix \ref{sec:sine}) shows a coherent signal from PKS J1309+1154 in all three light curves . The periods determined by the SWF method for the UMRAO, OVRO, and UMRAO+OVRO light curves are given in Appendix \ref{sec:sine}, where we tabulate them separately. The WWZ spectrum  (Fig.~\ref{plt:lightcurve1}(d)) clearly picks out a single period that dominates over the entire UMRAO+OVRO light curve.  None of the other 82 blazars showed a coherent peak over the duration of the observations in the WWZ spectrum.

The results of our  SWF analysis  shown in Table~\ref{tab:sine} of Appendix \ref{sec:sine}  provide a direct quantitative test of the periodicities in the 83 blazars in our sample. Some of these show the same period in the UMRAO and OVRO data, within the errors, but have periods  greater than 30 years, and large uncertainties. These periodicities are clearly not to be trusted both because of their large uncertainties and because spurious periodicities approaching the length of the observations are very common in blazar light curves.  Apart from those cases, 
\textit{only\/} PKS J1309+1154 has the same period within the errors (to $1.5\sigma$), with high fractional amplitude in the UMRAO and OVRO  light curves, and hence shows strong coherent sinusoidal variations throughout the 46-year observing period. Thus,  for the purposes of identifying potential SMBHB candidates, we consider \textit{only} PKS J1309+1154 as an SMBHB candidate, possibly showing the same characteristics as those found in PKS J0805--0111 and PKS 2131--021, discussed in Papers 1-4. 

In Papers 1 and 3 we described a method for simulating the light curve of any AGN of particular interest, based on the observed light curve, in which both the PSD and the  PDF of the flux-density variations are matched. The simulated light curve is for a gaussian process with  a power-law power spectrum  with the same index as observed. This makes it possible to generate  large numbers of matching simulated  light curves, and hence to make a reliable estimate of the significance of any feature in the observed light curve compared to random fluctuations  taking into account the steepness of the PSD and also taking into account the PDF.

It is important to note that in our estimates of significance we take into account the ``look elsewhere'' effect (Paper 1). In other words, since we have no \textit{a priori} reason for picking a particular period, we take into account all of the periods in the all of the simulations of a particular blazar, and not just the observed period. We call the corresponding significance the  \textit{global} significance.

In testing for real periodicities we set a threshold for the global significance at $p$-value=$ 1.35 \times 10^{-3}$ (i.e.,  $3\sigma$), and we regard only objects that pass this threshold as \textit{strong} SMBHB candidates.  PKS J1309+1154 is not a strong SMBHB candidate by this measure, since it has a global $p$-value=$ 1.06 \times 10^{-2}$ (i.e.  $2.3\sigma$).\color{black}

\begin{figure}[htp!]
    \centering
    \includegraphics[width=0.9\linewidth]{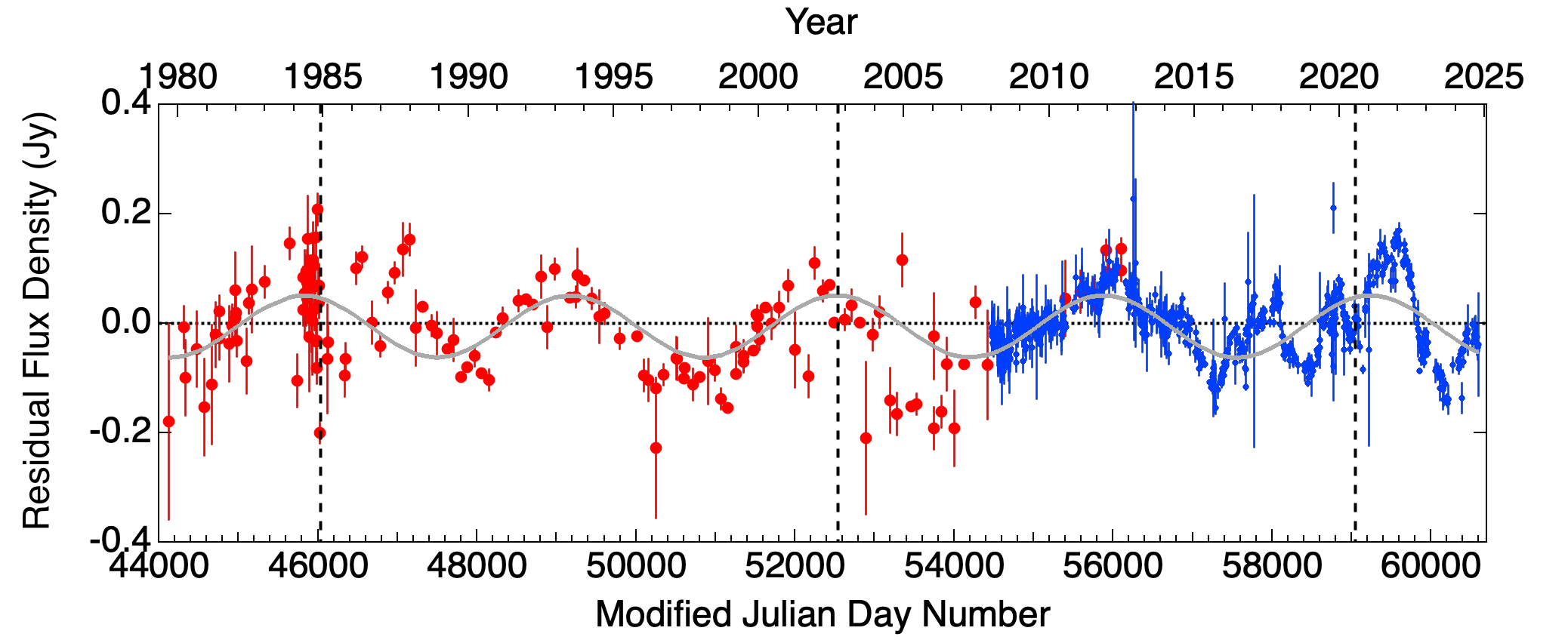}
    \caption{Residual light curve of PKS J1309+1154. The light gray curve is the least squares sine wave fit to the residual light curve. The vertical dashed lines indicate the times of the maxima in the sinusoidal curve in Fig.~\ref{plt:lightcurve1}(a).
    }
     \label{fig:residuals}
\end{figure}

\subsection{A Hint of a harmonic in PKS J1309+1154}\label{sec:hint}

The compelling detection of a harmonic in a blazar light curve would be an important development, since it would confirm that the fundamental is not simply a product of the steep PSD spectrum.  Harmonics are expected on the kinetic-orbital model of Paper 2, but, as discussed in that paper,  these are significantly smaller than those we will be considering here.

The fitted sine wave in Fig.~\ref{plt:lightcurve1}(a) has period $P=17.9$ yr and amplitude 0.22 Jy. The residual light curve of PKS J1309+1154 after subtraction of the sine wave plus linear trend,  shown by the black curve in Fig.~\ref{plt:lightcurve1}(a), is shown in Fig.~\ref{fig:residuals}.  This reduced the rms scatter in the light curve by a factor 2.9, from 0.178 Jy to 0.062 Jy. 

There is a hint of a periodicity in the residual light curve shown in Fig.~\ref{fig:residuals}, and the least squares fitted sine wave to this residual light curve has a period of 9.2 yr, which differs by only 3\% from the first harmonic of the 17.9 year period  at 8.95 yr. Note that this ``harmonic'' is the second strongest peak in the GLS periodogram shown in Fig.~\ref{plt:lightcurve1}(b). Subtraction of this subdominant sinusoidal component reduces the rms scatter in the residual light curve by a further 24\%, bringing the rms down by a factor 3.6 relative to that in the original light curve of Fig.~\ref{plt:lightcurve1}(a). 

The amplitude of this first possible harmonic is ($26\pm2$)\% of the amplitude of the fundamental.  In the case of PKS J2131, there is a marginal detection of a first harmonic of amplitude ($2\pm1$)\% of the fundamental (Paper 2).

Thus PKS J1309+1154 presents a case in which there may be a harmonic in the light curve detectable by successive subtractions of least squares sine wave fits.  This harmonic is very nearly in phase with the fundamental, the peaks of which are marked by the vertical dashed lines in Fig.~\ref{fig:residuals}.

We caution, however, that at this point the evidence for a harmonic is not strong because the sinusoidal variation with a period of 17.9 yr that we have subtracted from the original light curve introduces a periodicity into the light curve such that a fortuitous occurrence of random variations could then mimic a harmonic.  At this point, therefore, we regard this as a hint of a harmonic that should be borne in mind, together with the phase relationship, when thinking of physical models for this source. 

The confirmation of the harmonic would strengthen the case for PKS J1309+1154 being an SMBHB candidate, which has the advantage that it might possibly be confirmed in 9 yr rather than the 18 yr needed to confirm the reality of the fundamental.

Note that the approach we have adopted above, namely of selecting the most dominant sinusoidal component by least squares fitting, subtracting it from the light curve, and then selecting the next most dominant component, and fitting it by least squares and subtracting it, is deliberate.  We believe this to be preferable to fitting two components simultaneously because it is important to confirm by visual inspection that, at each stage, there is clear evidence of  another periodicity in the residual light curve. This approach of successively subtracting a feature and then examining the residual for the next most prominent feature,  is the same as that used with great success in image restoration using the CLEAN method  \citep[e.g.,][]{1984ARA&A..22...97P} in which successive  components are subtracted from a ``dirty'' image only when there is clear evidence for that component in the residual map.

\section{Other Observations  of PKS J1309+1154}\label{sec:j1309}

In this section we describe other observations of PKS J1309+1154 and show that there are several similarities with PKS 2131-021, which, while not definitive, are certainly suggestive of a common cause of the periodicities seen in these two objects.

\subsection{VLBA observations of PKS J1309+1154}\label{sec:vlbsa}

\begin{figure}[htp!]
    \centering
    \includegraphics[width=0.8\linewidth]{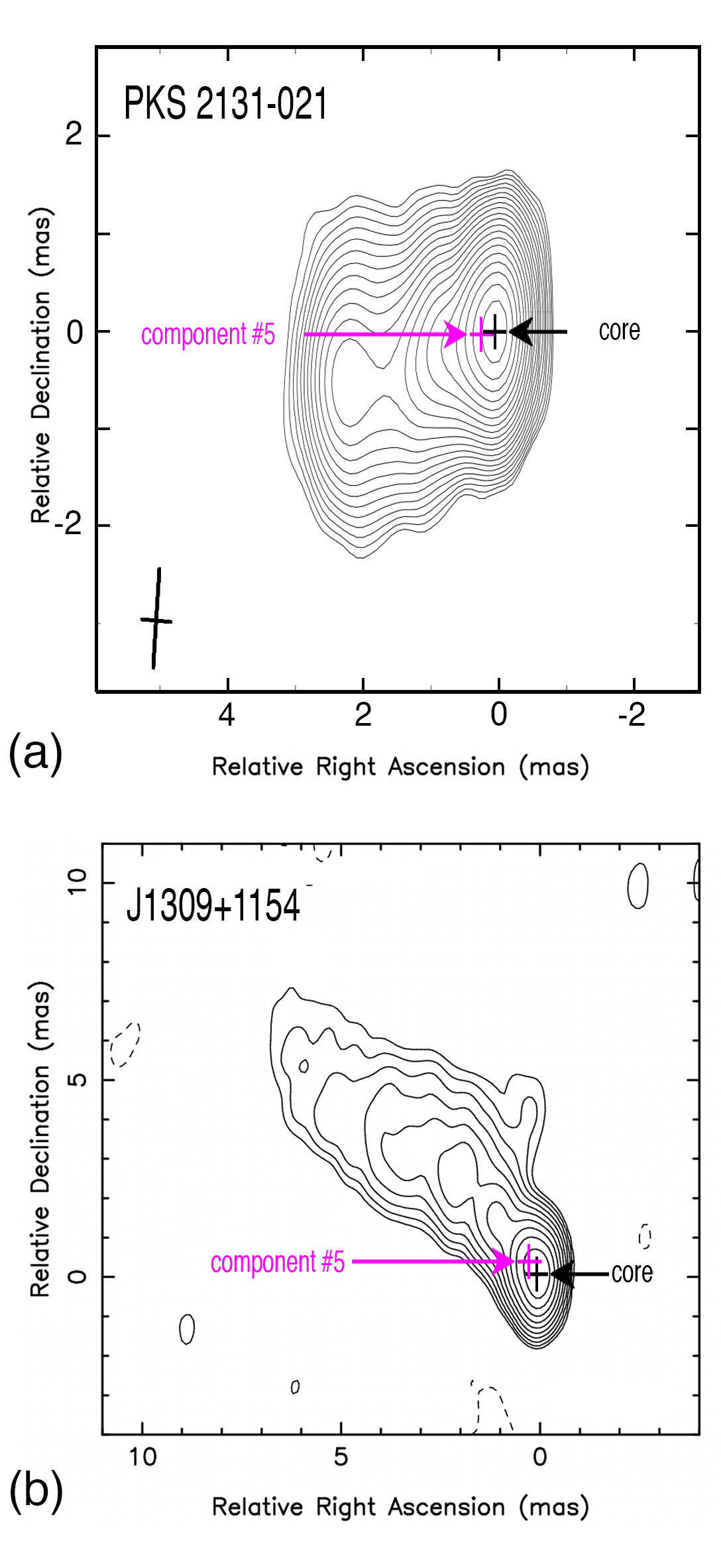}
    \caption{VLBA 15 GHz MOJAVE maps of (a) PKS 2131--021, and (b) PKS J1309+1154, showing the close juxtaposition of the stationary component \#5 and the core. In both cases the sinusoidal variations originate not only in  the core, but also in component \#5.}
    \label{fig:vlbacomp5}
\end{figure}

The Monitoring of Jets in Active Galactic Nuclei with VLBA Experiments (MOJAVE) program \citep{2021ApJ...923...30L}  monitors bright radio-loud active galactic nuclei (AGN)  using high-resolution Very Long Baseline Array (VLBA) observations at 15 GHz.  Fortunately, PKS J1309+1154 has been observed 13 times on  the MOJAVE program. We undertook an analysis of the MOJAVE results on PKS J1309+1154  to determine whether the observed long-term periodic modulation originates in the core or the jet.

\begin{figure}[htp!]
    \centering
    \includegraphics[width=\linewidth]{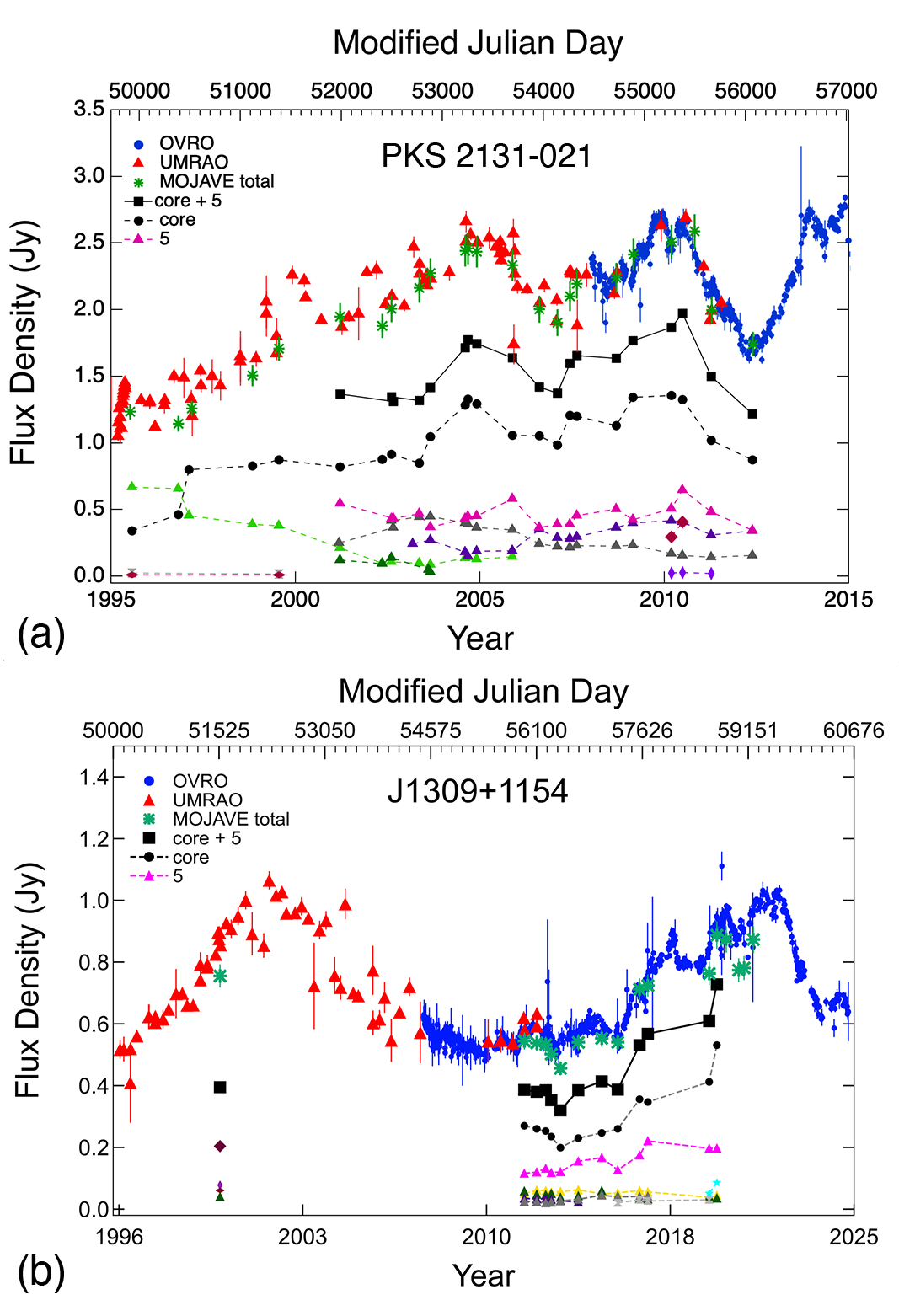}
    \caption{UMRAO+OVRO monitoring and MOJAVE VLBA results on PKS J1309+1154 and PKS 2131--021. (a) PKS 2131--021 adapted from Paper 1. (b) PKS J1309+1154 from MOJAVE.  In both cases the combined flux densities of the core component and component \#5 follow the total flux density and are clearly the major contributors to the sinusoidal variation.  It is just coincidental that in both sources the stationary component is \#5. }
    \label{fig:vlba}
\end{figure}

The total flux densities of the core and the jet, at  a number of epochs, are provided by  MOJAVE as part of their long-term monitoring program. The  corresponding published model-fitted components are available on the VizieR catalogue service. These data come from the supplementary material of Lister et al. (2021). The component models were derived by Lister and collaborators using Gaussian \color{black} model fits to the \color{black} VLBA visibility curves with the DIFMAP software \citep{1994BAAS...26..987S,1997ASPC..125...77S}. The entry for each component includes the peak flux density, the radial separation from the core, the position angle, the FWHM size of the component's major  axis, and its axial ratio. We retrieved the component  data for multiple epochs. There is one epoch from 1999 and a  series of epochs spanning 2011 to 2019. 

\subsection{The Stationary Component}\label{sec:stationary}

Of the 83 blazars in the combined UMRAO+OVRO sample, in 40\% of them the components within a beam width of the core are stationary. The positions of component \#5 relative to the core for the MOJAVE observations of  PKS J1309+1154 and PKS 2131--021 are shown in Fig.~\ref{fig:vlbacomp5} and listed in Table \ref{tab:stationary}. 

We see  that the position of component \#5 relative to the core can be measured with high precision, and that it is unchanging in both blazars. In VLBI it is possible to measure the distance between components with an uncertainty equal to the beamwidth divided by the signal-to-noise ratio, when the signal-to-noise ratio is high as it is here, by model fitting in the $(u,v)$ plane.

\begin{deluxetable*}{c@{\hskip 8mm}ccccc}
\tablecaption{The Position of Component \#5 Relative to the Core in the Jets of PKS J1309+1154 and PKS 2131--021}
\tablehead{PKS J1309+1154&PKS J1309+1154&PKS J1309+1154&PKS 2131--021&PKS 2131--021&PKS 2131--021\\
Date&Separation (mas)&Position Angle &Date&Separation (mas)&Position Angle}
\startdata
12/29/11	&	0.48	&	$35.8^\circ$	&	3/15/01	&	0.29	&	$85.7^\circ$	\\
6/25/12	&	0.46	&	$36.5^\circ$	&	8/28/02	&	0.36	&	$91.1^\circ$	\\
11/2/12	&	0.47	&	$37^\circ$	&	8/7/02	&	0.36	&	$93.2^\circ$	\\
1/21/13	&	0.48	&	$38.5^\circ$	&	5/9/03	&	0.29	&	$93.4^\circ$	\\
6/2/13	&	0.44	&	$40.7^\circ$	&	8/28/03	&	0.32	&	$96.0^\circ$	\\
2/14/14	&	0.49	&	$41.1^\circ$	&	8/9/04	&	0.33	&	$91.8^\circ$	\\
1/18/15	&	0.51	&	$39^\circ$	&	9/2/04	&	0.33	&	$94.9^\circ$	\\
9/6/15	&	0.49	&	$38.8^\circ$	&	12/2/04	&	0.30	&	$87.0^\circ$	\\
7/16/16	&	0.46	&	$34.9^\circ$	&	11/17/05	&	0.33	&	$93.3^\circ$	\\
11/12/16	&	0.47	&	$33.8^\circ$	&	8/9/06	&	0.34	&	$93.0^\circ$	\\
4/15/19	&	0.47	&	$35.6^\circ$	&	2/5/07	&	0.3	&	$93.4^\circ$	\\
8/4/19	&	0.49	&	$35.9^\circ$	&	6/10/07	&	0.28	&	$88.4^\circ$	\\
	&		&		&	8/16/07	&	0.27	&	$88.3^\circ$	\\
	&		&		&	9/12/08	&	0.31	&	$91.6^\circ$	\\
	&		&		&	2/25/09	&	0.31	&	$90.1^\circ$	\\
	&		&		&	3/10/10	&	0.31	&	$92.9^\circ$	\\
	&		&		&	6/27/10	&	0.30	&	$94.1^\circ$	\\
	&		&		&	4/11/11	&	0.30	&	$91.7^\circ$	\\
	&		&		&	5/24/12	&	0.30	&	$91.6^\circ$	\\
    \hline
mean	&	$0.476$	&	$37.3^\circ$	&	mean	&	$0.312$	&	$91.7^\circ$	\\
standard deviation	&	$\pm0.018$	&	$\pm2.3^\circ$	&	standard deviation	&	$\pm0.024$	&	$\pm2.7^\circ$	\\
std error in mean	&	$\pm0.006$	&	$\pm0.7^\circ$	&	std error in mean	&	$\pm0.006$	&	$\pm0.6^\circ$	\\
\enddata
\tablecomments{The standard deviations indicate the rms scatter in the measurements, and show that these values repeat with high precision from epoch to epoch, so that the standard error in the mean is very small, and consequently there is no doubt that component \#5 is stationary relative to the core to within these uncertainties in both PKS J1309+1154 and PKS 2131--021.  These values are all from the MOJAVE program.} 
\label{tab:stationary}
\end{deluxetable*}

\begin{figure}[hbp!]
    \centering
    \includegraphics[width=\linewidth]{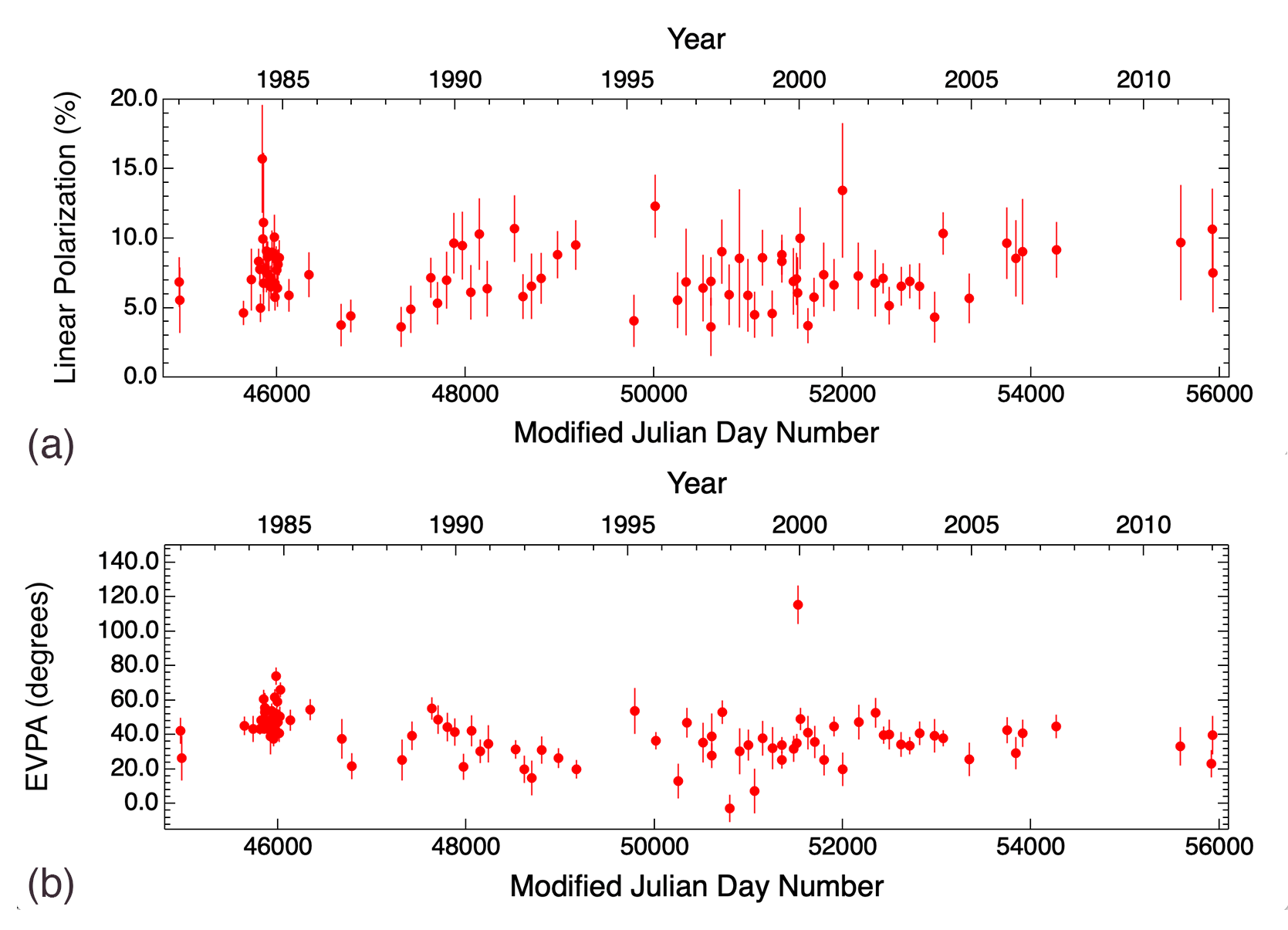}
    \caption{The polarization of PKS J1309+1154 at 14.5 GHz from the UMRAO monitoring survey. (a)  the fractional linear polarization. (b) the electric vector position angle (EVPA). The mean EVPA  is aligned with the inner jet axis.}
    \label{fig:poln}
\end{figure}

In Fig.~\ref{fig:vlba} we show the flux densities of the key components together with the UMRAO+OVRO light curve showing the total flux densities over the relevant epochs.  

Since the sinusoidal variations in total flux density dominate the light curve, it is easy to determine which components in  the VLBA decompositions contribute to this variation.  While we do not have VLBA maps throughout the  whole period monitored, we do have maps over two cycles of the sinusoid in PKS 2131--021 and over half a cycle of the sinusoid in the case of PKS J1309+1154.  As can be seen in Fig.~\ref{fig:vlba}, the combined flux densities of the core plus component \#5 dominate the flux density variations.   It is clear, therefore, that the sinusoidal variations in total flux density seen in these two objects originate in these two components, i.e., the cores and components \#5.

\subsubsection{Possible Explanation for the Stationary Components}\label{sec:explanation}

 Stationary components in the jets of blazars have been much discussed in the literature \citep[e.g.,][]{2014ApJ...787..151C,2015ApJ...803....3C,2020AandA...640A..62A,2022ApJS..260...12W,2024AandA...692A.127A,2025arXiv250612457A}, and are usually ascribed to standing shocks in the jet. Here we offer an alternative explanation.

In the context of a binary black hole interpretation, sinusoidal variation of blazar flux density is associated with orbital motion of the jet source with period $P$. The simplest models suppose that the direction of the jet launch velocity is modulated by the orbital velocity. The variation is determined more by the changing direction than the changing speed. However, as the jets are directed at us ultrarelativistically, the emission seen at a given observer time may originate from radii $\gg cP$. This may be problematic for explaining the variation.

An alternative model posits that the jet direction is determined by the orbital motion of the confining medium, most reasonably, though not necessarily, a non-relativistic MHD wind. The jet keeps an ultrarelativistic speed within a channel. In this case, the variable emission comes from a range of radii with each frequency's minimum radius separated by $\lesssim cP$ (Sullivan et al.\ 2025, in preparation). If this model is relevant, then it is tempting to identify the ``stationary'' jet component seen in PKS J1309+1154 and PKS 2131$-$021, and perhaps other sources, with the outer radius of the confining wind before the interaction is dominated by the surrounding interstellar medium which, presumably does not share the orbital motion. It is natural to associate this with a Bondi or recollimation radius \citep[e.g.,][]{2019ARAandA..57..467B},  which could exhibit flux modulation without proper motion. The actual radius of this component could be $\gtrsim cP$. Future magnetohydrodynamic simulations of orbiting jets, disks and winds should be instructive.

\begin{figure}[htp!]
    \centering
    \includegraphics[width=0.8\linewidth]{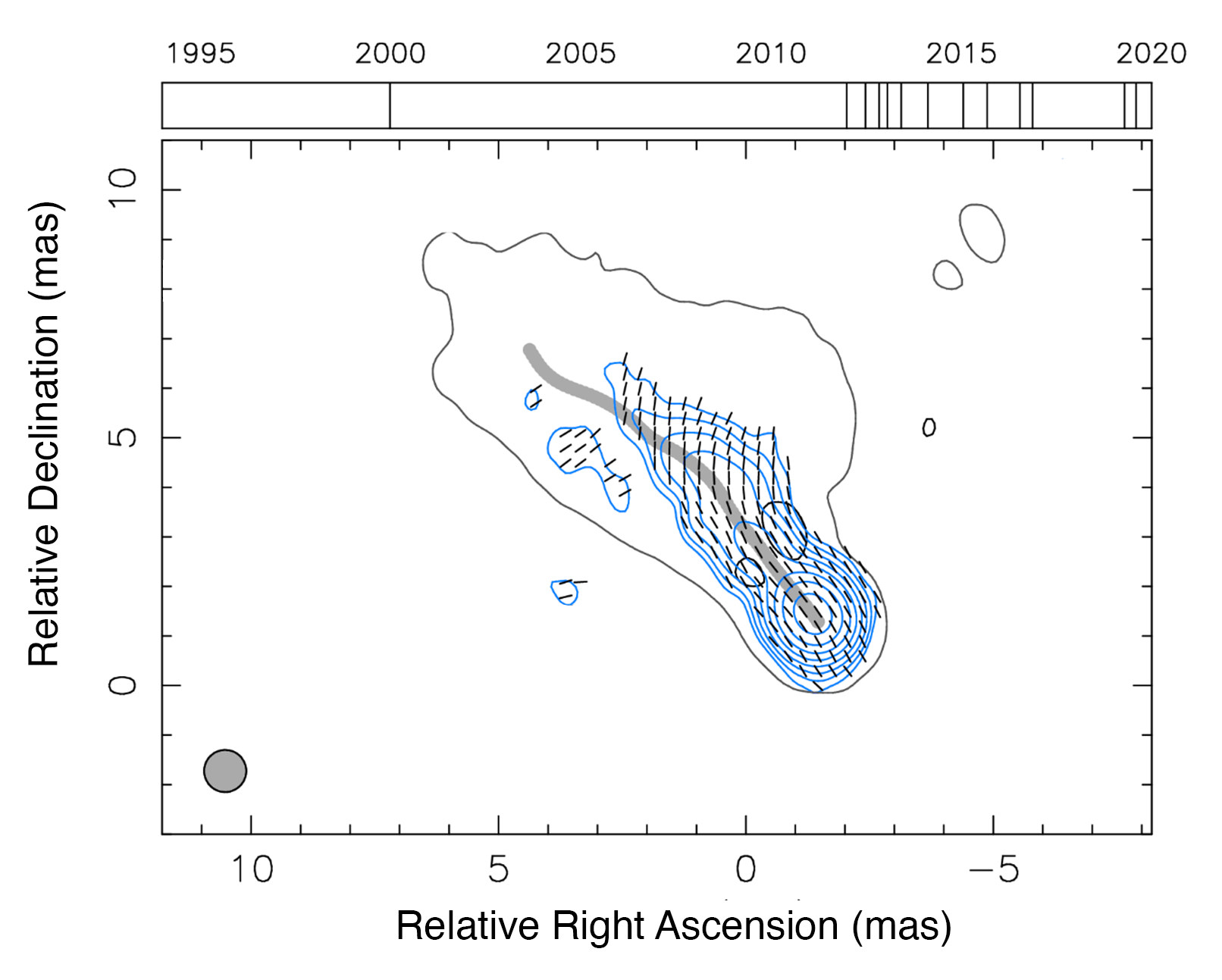}
    \caption{Stacked MOJAVE polarization map of PKS J1309+1154 for the 13 epochs that it was observed, adapted from \citet{2023MNRAS.520.6053P}. The vertical bars above the map indicate the epochs of observation. The gray line indicates the ridge line of the jet. The EVPA vectors are indicated by the short black lines. }
    \label{fig:polnmap}
\end{figure}

\subsection{Polarization of PKS J1309+1154 at 14.5 GHz}\label{sec:poln}

The total linear polarization of PKS J1309+1154 was monitored at the UMRAO at 14.5 GHz for the duration of the light curve, and is shown in Fig. \ref{fig:poln}.  The mean electric vector position angle (EVPA), which is dominated by the core and inner jet,  is $39.1^\circ\pm 1.7^\circ$, which may be compared with the position angle of component \#5 relative to the core given in Table. \ref{tab:stationary}, of $37.3^\circ \pm 0.7^\circ$. Thus the difference between the position angle of the inner jet and the EVPA is $1.8^\circ \pm 1.9^\circ$.  This shows that the magnetic field is perpendicular to the jet axis, i.e. it is a helical magnetic field, as is also the case in PKS 2131-021. The stacked MOJAVE polarization map of PKS J1309+1154 is shown in Fig. \ref{fig:polnmap}.

\subsection{ALMA Observations of PKS J1309+1154}\label{sec:J1309alma}

We downloaded ALMA data from the ALMA calibrator website. ALMA has observed PKS J1309+1154 a number of times in Band 3  (at 91.5 GHz and 104 GHz). Since the data are few and almost all the observations were made at both frequencies, we make no distinction, but  refer to them as ``Band 3''.

\begin{figure}[htp!]
    \centering
    \includegraphics[width=\linewidth]{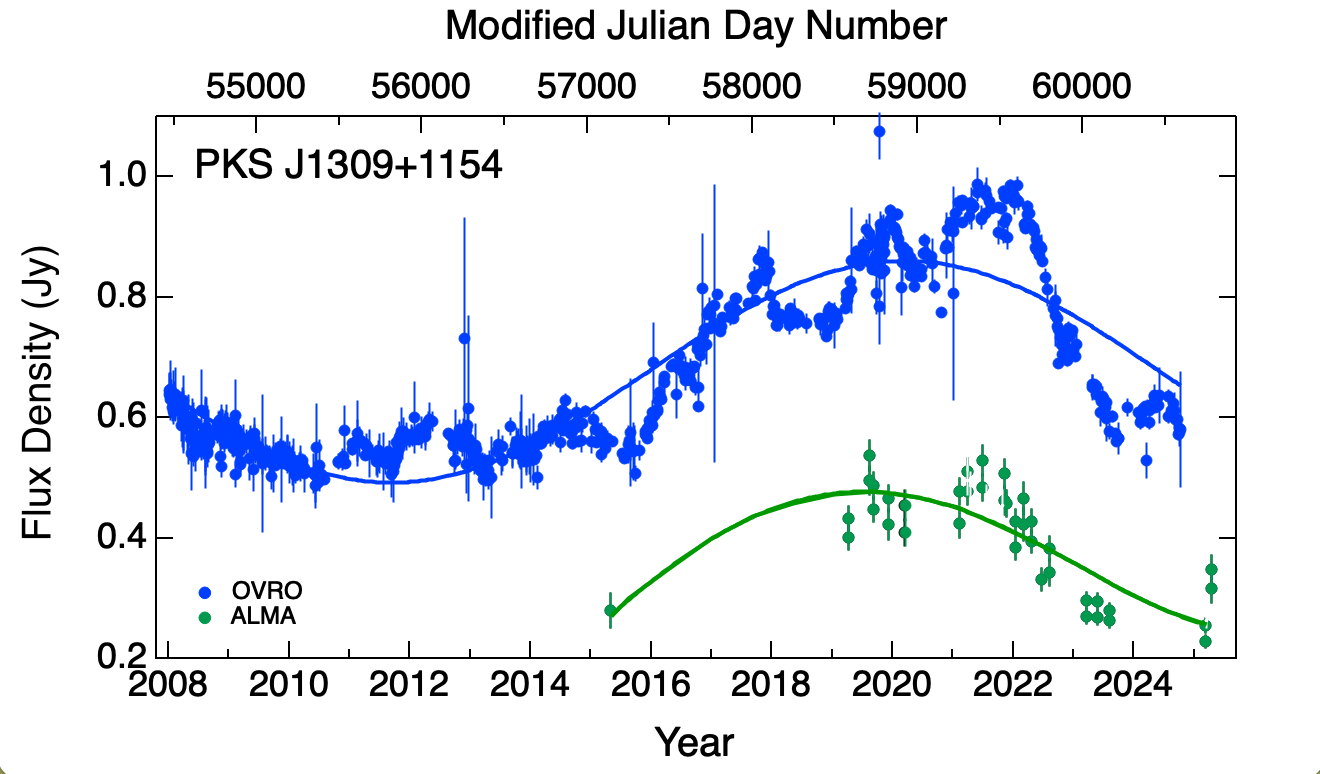}
    \caption{The OVRO (blue)  and ALMA (green) light curves of PKS J1309+1154, together with their least squares sine wave fits.   }
    \label{fig:alma}
\end{figure}

 In Fig.~\ref{fig:alma} we show the ALMA Band 3 and OVRO light curves of PKS J1309+1154. As discussed in detail in Paper 2, when comparing sinusoidal variations at different frequencies it is important to  fit the sine wave over the window of the observations.  This is because, as discussed in detail in Paper 1, random correlated variations in the non-sinusoidally varying components cause apparent variations in the period of the sinusoid of up to 10\%.  For this reason, we do not use the period of the whole UMRAO+OVRO light curve in this comparison with the ALMA data, but only that of the OVRO data. 
 
 We have carried out a least squares fit  to  the OVRO light curve of Fig.~\ref{fig:alma}, and we find a period of  6201 days (16.5 yr), i.e., 8\% shorter than the period fitted over the full 46 years.  We then held this period fixed and  fitted the sine wave to the ALMA data shown in  Fig.~\ref{fig:alma}.  The two sine waves are offset in phase by $-300^{+260}_{-200}$ days, with the higher frequency observations leading the lower, although the uncertainties are large.  This same phenomenon has been seen in both PKS 2131$-$021 and PKS J0805$-$0111. In the case of PKS 2131$-$021, the ALMA  Band 3 variations lead the OVRO 15 GHz variations by 0.075 of a period (Paper 2). In the case of PKS J0805$-$0111, the ACT 95 GHz variations lead the OVRO observations by 0.13 of a period (Paper 4).  In this case we see that the ALMA Band 3 sinusoidal variation leads that of OVRO at 15 GHz by $0.05^{+0.4}_{-0.3}$ of a period.
 
 The similarity,  with respect to the small phase shift,  between PKS J1309+1154 and the  two strong SMBHB candidates,  PKS 2131$-$021 and PKS J0805$-$0111, provides more evidence suggesting that PKS J1309+1154 is a good, albeit not a strong, SMBHB candidate.

\begin{figure*}[htp!]
    \centering
    \includegraphics[width=0.9\linewidth]{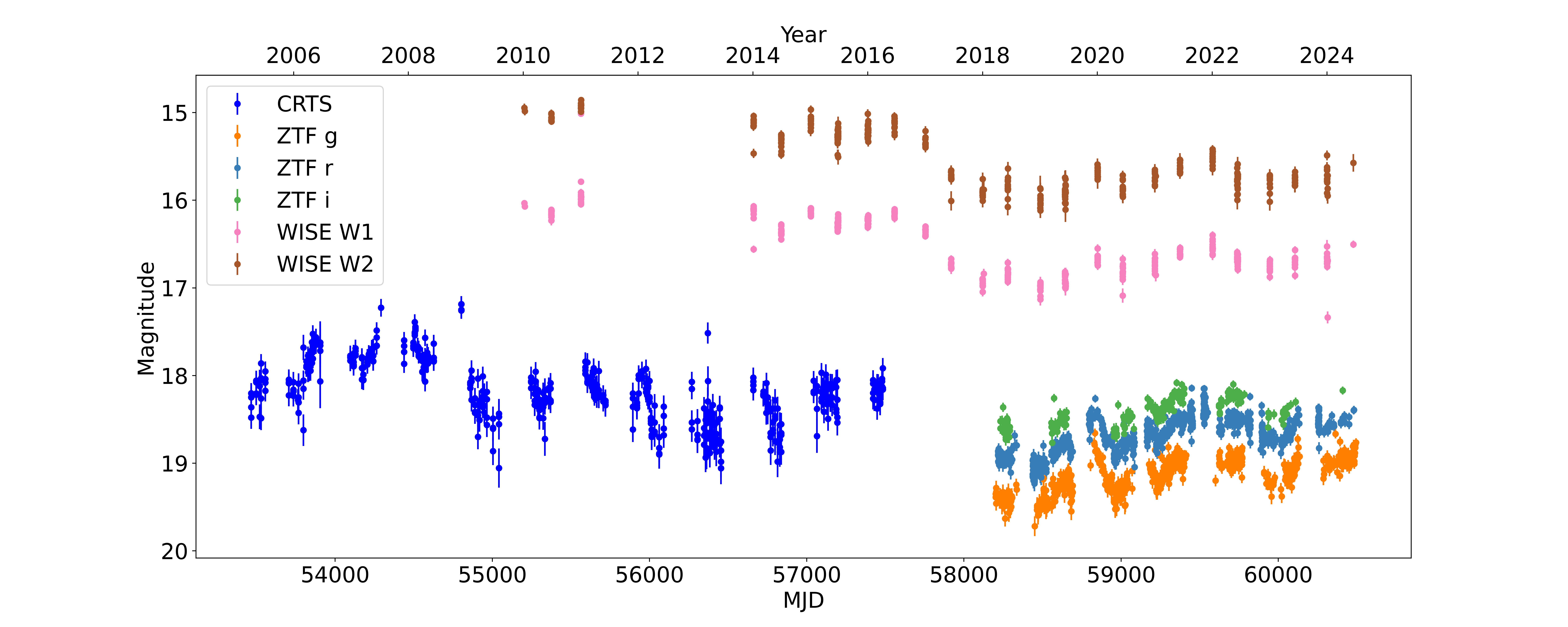}
    \caption{Optical (CRTS and ZTF \textit{gri}-band) and NIR (WISE W1 and W2-band) light curves for PKS J1309+1154. W1 and W2 magnitudes are shifted up by 3 mags for clarity.}
    \label{fig:oir_dat}
\end{figure*}

\subsection{Optical and Infrared Observations of PKS J1309+1154}\label{sec:J1309optical}

Before discussing the optical and infrared properties  of PKS J1309+1154, we first consider those of the SMBHB candidate PKS 2131--021. PKS 2131--021 has recently been shown to have coherent sinusoidal variations from radio to the optical wavelengths, exhibiting a strong correlation between phase and the observed frequency. According to the kinetic-orbital model \citep{2017MNRAS.465..161S,2025ApJ...985...59K}, this phase shift is due to the fact that the higher-frequency emission originates farther down the blazar jet axis closer to the binary. 

We have therefore assembled  optical and infrared observations of PKS J1309+1154, to determine whether the radio  sinusoidal variations can also be seen in other wavebands.

PKS J1309+1154 has been monitored at optical wavelengths spanning about 19.25 years (MJD 53500--MJD 60500) by the Catalina Real Time Transient Survey \citep[CRTS]{Drake2009} and later the Zwicky Transient Facility \citep[ZTF]{2019PASP..131a8002B} with a small gap in coverage between the two surveys of $\sim800$ days. In Figure \ref{fig:oir_dat}, we show the optical light curves, noting that CRTS has no standard filter. We discard a single erroneous measurement that is 2 magnitudes brighter than the rest of the data. We appended multi-epoch ALLWISE and NEOWISE W1 and W2-band photometry \citep{2011ApJ...731...53M, 2019ipac.data...I1W}, shifted up by 3 magnitudes for easier comparison to the optical photometry.

While they show some modulation, the optical light curves do not show a clear modulation with the period detected by the UMRAO and OVRO light curves. Visually, there appears to be a maximum around 2007--2008 in the CRTS light curve, and the ZTF data might be rising towards a maximum either around 2022 or after 2024 despite when accounting for an overall decreasing trend in the combined optical light curves.

If we combine the CRTS light curve (calibrated to ZTF r) with the ZTF r-band light curve and convert to fluxes, a GLS analysis reveals a signal at 24 years with a GLS power of $\mathcal{P}$=0.50.\footnote{To distinguish the GLS power from the period of the periodicity, $P$,  we use the symbol $\mathcal{P}$ to denote the GLS power. }Alternatively, given that the r-band data appear to exhibit a slow decreasing trend, de-trending the r-band light curves using a linear fit results in two signals at ${\sim} 7.5$ years with power $\mathcal{P}=0.42$ and ${\sim}17.1$ years with power $\mathcal{P}=0.37$. Finally, combining all optical data, a multi-band LS analysis using \textsc{astropy}'s \textsc{LombScargleMultiband} (based on \citealt{2015ApJ...812...18V}) reveals a signal at 11 years with a power of $\mathcal{P}=0.47$. 
These signals are relatively weaker and individually inconsistent with the radio period of $17.9$ years; a coherent sinusoidal variation cannot be concluded with the current optical data. However, these signals do bracket the radio period, and we note that the phase of the CRTS peak at MJD $52473$ under a period of 17.9 is ${\sim}-0.3$. This is broadly consistent with the NIR and optical phase shifts observed in PKS 2131--021 (being $-$0.27 and $-$0.35, respectively). Therefore, the possibility of coherent variation in the optical should not be ruled out without a larger baseline of observations.




We note that two redshifts have been cited for PKS J1309+1154 in the literature: a photometric redshift of $z=0.415$ from \citep[SDSS DR6]{2009ApJS..180...67R} and later a spectroscopic $z=1.415$ from \citep[SDSS DR13]{2017ApJS..233...25A}. However, the SDSS BOSS spectrum reveals an effectively featureless optical spectrum typical of blazars\footnote{\url{https://specdash.idies.jhu.edu/?catalog=sdss&specid=6104739071615326208}}, leading us to conclude that no reliable redshift estimate can be determined. At least two absorption systems are evident in the BOSS spectrum, at redshifts of 0.515 and 0.852, traced primarily by the Mg\,II $\lambda$ $\lambda$2796, 2803 doublet. Thus, PKS J1309+1154 is at a redshift $z\geq 0.852$. 

\subsection{X-ray Observations of PKS J1309+1154}\label{sec:1309xray}

PKS J1309+1154 was observed by \textit{Swift-XRT} during the past decades. We extracted and combined legacy X-ray spectra using the archival data on the UK Swift Science Data Centre (UKSSDC). The final combined spectrum has effective exposure time of 8486\,s in the$0.3-10\,\rm keV$ band. The combined spectrum is binned to have at least 1 count/bin and avoid empty channels (with \textsc{ftgrouppha}). The final spectrum has 339 bins.

Models were fitted to the X-ray spectrum  using \textsc{Xspec} \citep{Arnaud1996}. During the fitting, the modified Cash statistic includedGalactic foreground absorption of $1.9\times10^{20}\rm\,cm^{-2}$, we find that the X-ray spectrum can be well fitted with a power-law of a photon index of $\Gamma=1.28\pm0.26$ (C-stat/DoF = 227.2/337), significantly harder than other blazars that show strong evidence of sinusoidal variation ($\Gamma\simeq1.8-2.2$ for PKS 2131--021 and PKS 0805--011). We estimate the total flux in $0.3-10\,\rm keV$ from the best-fit model to be $7.5\times10^{-13}\,\rm erg~s^{-1}~cm^{-2}$ (in the observer's frame). Simulations have predicted that close-separation accreting SMBHBs may have a periodically modulated hard X-ray component whose period is around the binary orbital period \citep{Tang2018MNRAS,Krolik2019ApJ}.  Future X-ray timing studies with large field-of-view X-ray instruments (e.g., \textit{Swift} and \textit{eRosita}) could help unveil the nature of this target. 

\section{``Harmonics'' In Blazar Light Curves}\label{sec:harmonics}

A number of blazars have been reported as showing multiple periodicities, some of which are harmonics,  in their radio, optical, or gamma-ray light curves \citep[e.g.,][]{2000ApJ...545..758R,2006ApJ...650..749L,2013MNRAS.434.3487A,2014MNRAS.443...58W,2017ApJ...847....7B,2021MNRAS.501.5997T,2022RAA....22k5001D,2023MNRAS.523L..52B,2023arXiv231212623S,2024MNRAS.531.3927M,2025A&A...698A.265L}.

It is important to note at the outset of our discussion of the harmonics in these 83 blazars that in the case of the SMBHB candidates PKS J1309+1154 and PKS 2131--021 the harmonics expected on the kinetic-orbital model are very small, and of no relevance here (Paper 2).

We find that we can have confidence in the spectral peaks identified by these methods only in cases where the PSD spectra are dominated by a single long-lived sinusoidal component, such as is the case, for example, in PKS 2131--021 (Papers 1, and 2), in PKS J0805--0111 (Paper 3), and in PKS J1309+1154 (Fig.~\ref{plt:lightcurve1}). In the last case only $\sim 2.5$ cycles have been observed, so the consistent GLS and WWZ spectra and the good sinusoidal fit shown in Fig.~\ref{plt:lightcurve1} could well be a transient characteristic of this source. Among our 83 sources there are many which, for short durations, exhibit similar features.

Although the standard GLS analysis method can detect multiple peaks indicative of multiple sinusoidal periodicities in blazar light curves, the model behind it is a single sinusoid embedded in white noise \citep{1982ApJ...263..835S,2003sca..book..309B}.  Therefore, in principle the standard GLS model is inappropriate for a light curve with  two sinusoidal components.  However, in practice the standard GLS power spectrum reliably detects sinusoidal 
components if their frequencies are resolved -- that is, separated in frequency by more than the width of the spectral window function, and this is what we have used in this paper. At the end of this study we became aware of the extension of the  standard GLS model to the case of ``multi-harmonic'' periodograms \citep{2009MNRAS.395.1541B,2020arXiv200110200S} and we have started an investigation of this approach together with a machine learning approach.

\begin{figure}[htp!]
    \centering
    \includegraphics[width=\linewidth]{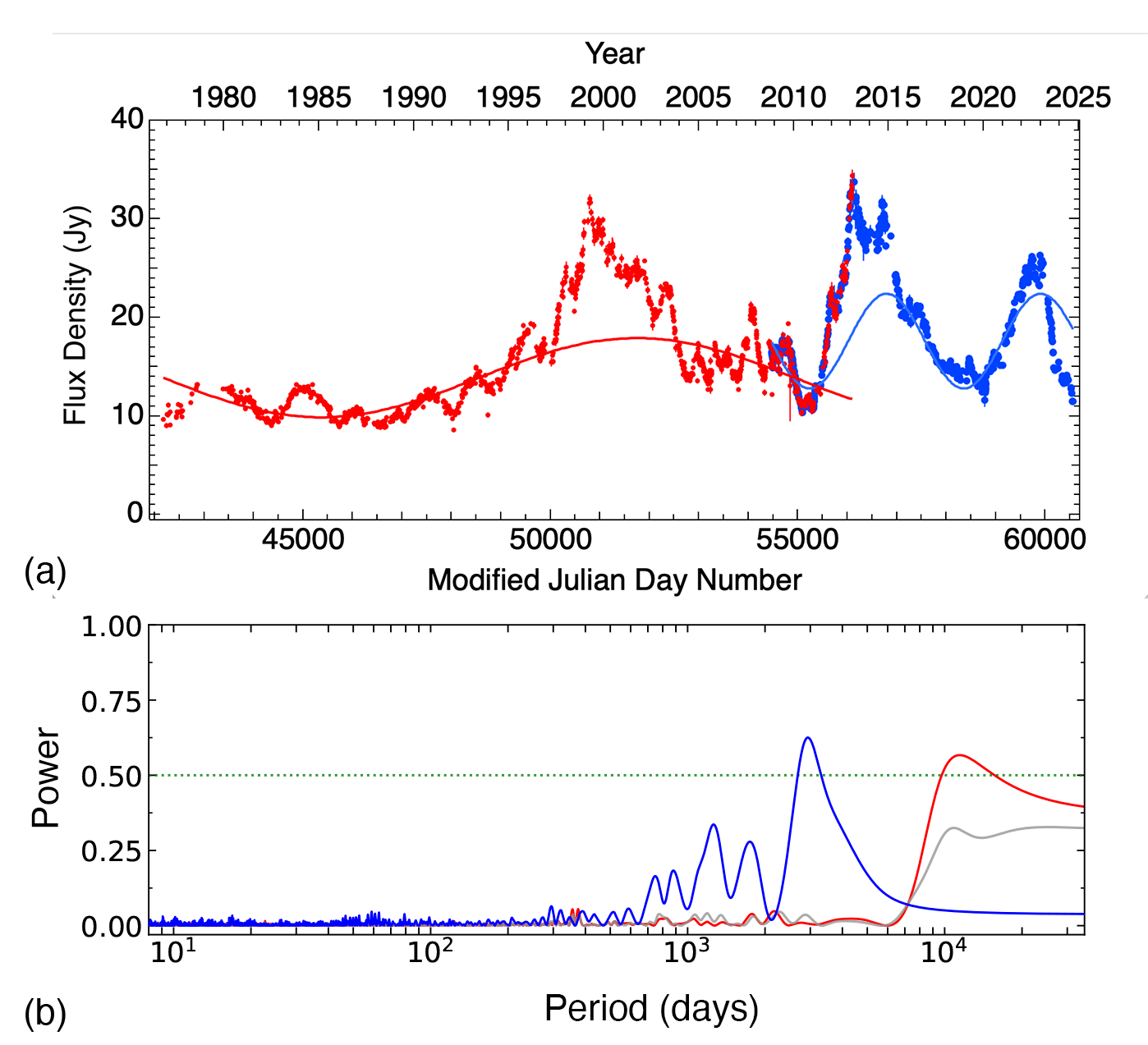}
    \caption{Light curves and GLS spectra of 3C 279. (a) the combined UMRAO 14.5 GHz and OVRO 15 GHz light curves (red points:UMRAO, blue points:OVRO).  The fitted sine waves have periods $P=12,788$ days and  $P=3,118$ days for UMRAO and OVRO, respectively (see Table \ref{tab:sine}).  (b) The GLS spectra of the UMRAO light curve (red), the OVRO light curve (blue), and the combined UMRAO+OVRO light curve (gray). The green dotted line marks the $\mathcal{P}$=50\% power level (see text).
    }
     \label{fig:3c279}
\end{figure}

Apart from PKS J1309+1154, the variability spectra of the other 82 blazars from 1979 to 2008 were markedly different from their counterparts between 2008 and 2025.   This demonstrates clearly that we are not observing a statistically stationary phenomenon in blazar light curves, even with a timespan of 16 years.

\begin{figure*}[htp!]
    \centering
    \includegraphics[width=1\linewidth]{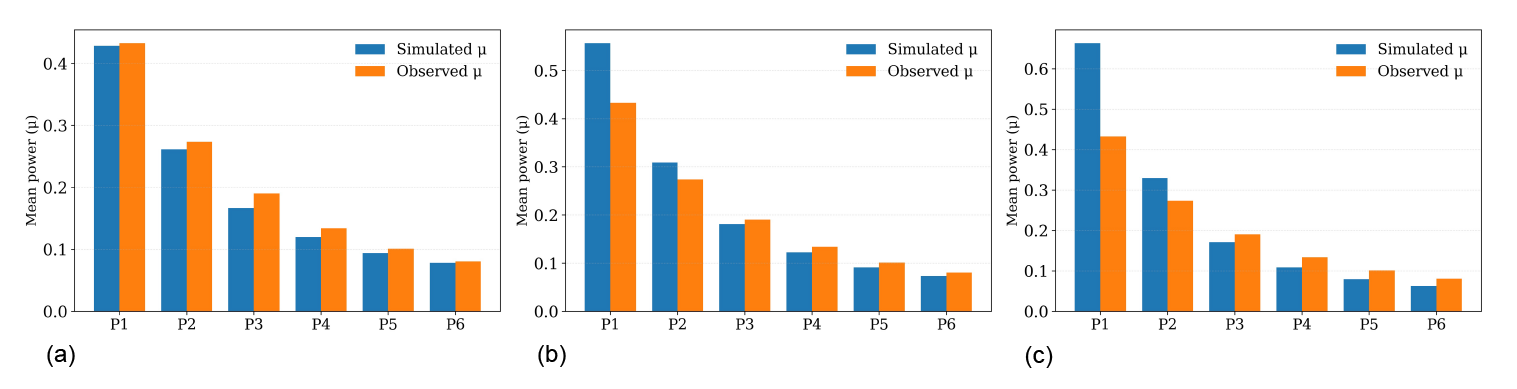}
    \caption{The mean powers of the six most powerful peaks in the GLS spectra of the simulations compared to the observations: (a) for $\beta\,=\,-1.5$, (b) for $\beta\,=\,-1.9$, and  (c) for $\beta\,=\,-2.5$.
    }
     \label{fig:betadistns}
\end{figure*}

\begin{deluxetable*}{l@{\hskip 8mm}cccc}
\tablecaption{Comparison of the Harmonics in the Observed  \textit{vs\/} Simulated Light Curves for $\beta\,=\,-1.5$  }
\tablehead{Category&Number of&Number of&Harmonics&Ratio of Observed to\\
&Light Curves&Harmonics&per source&simulated Harmonics per source}
\startdata
Observed&83&$371\pm19$&$4.47\pm0.23$&$1.17\pm0.06$\\
Simulated&8,300&$31,720\pm178$&$3.82\pm0.02$&\\
\enddata
\tablecomments{All uncertainties on $N$ are assumed to be $\sqrt{N}$. Numbers above are for the  case $\beta\,=\,-1.5$.  Almost identical results are obtained for $\beta\,=\,-1.9$, and $\beta\,=\,-2.5$.} 
\label{tab:harmonics}
\end{deluxetable*}

An illuminating example is that of 3C 279, shown in Fig.~\ref{fig:3c279}.  The UMRAO and OVRO light curves shown in  Fig.~\ref{fig:3c279}(a) look fairly similar, and they overlap for four years (2008--2012). Nevertheless the GLS spectra shown in \ref{fig:3c279}(b) are qualitatively very different.  The OVRO GLS spectrum exhibits a number of peaks with powers between 0.2 and 0.4, whereas the UMRAO spectrum exhibits no peaks having power greater than $\mathcal{P}=7\%$ apart from that associated with the longest period peak.  One might think that this difference arises partly because there might be more power in that peak, but its power is $\mathcal{P}=66\%$, i.e., not that different to the power in the largest OVRO peak ($\mathcal{P}=63\%$). Thus, even strong GLS peaks having power $\mathcal{P}>0.6$ that are a factor $\gtrsim 2$ stronger than the other peaks in the GLS spectrum are not indicative of true periodicities in the blazar.  There are, furthermore, many harmonic relationships between the periods of the peaks in both the UMRAO and the OVRO GLS spectra. The striking differences between the GLS spectra for UMRAO and OVRO, which are similar to those seen in many of our sources, show that these periodicities and their harmonics are not stable, reproducible features of these blazars. Note that, in the combined 50-yr UMRAO+OVRO GLS spectrum of 3C 279 shown in Fig.~\ref{fig:3c279}(b), apart from the two peaks with periods above $10^{4}$ days the largest peak has power of only $\mathcal{P}=5.1\%$, while the others are 4\% and lower. This shows that the power is spread across the variability spectrum even in the presence of large fractional flux-density variations.   This example shows that even for GLS peaks with power as high as $\mathcal{P}=60\%$, the main periodicity should be interpreted with caution.  Clearly, the much lower powers of the apparent harmonics should be interpreted with still more caution.

\subsection{Harmonics in simulated light curves}\label{sec:harmsim}

In this section we present a comparison between the apparent periodicities in the 83 blazars in our sample, as revealed through GLS periodogram analysis, with simulated light curves constructed as described above in Section~\ref{sec:qpos}. 

In order to test the reality of the apparent harmonics we see in our GLS spectra of all 83 blazars in our sample, we  generated 300 pure random noise simulations with no coherent sinusoids added, to match the cadence and noise characteristics of each of our 83 blazars.  We constructed 100 simulations for each of  the three different power law PSD slopes: $\beta = -1.5,\;\beta = -1.9$, and $\beta = -2.5 $ for each blazar.    We then carried out GLS spectral analysis of these 24,900 simulated light curves.

Note that the above values of $\beta$ span the range of PSD slopes seen in many blazars at optical wavelengths in which the variations may be characterized by a damped random walk  with a turnover from $\beta = -2$ to $\beta = 0$ at frequencies that can be below $10^{-3}\; \textrm{day}^{-1}$ \citep{2014ApJ...788...33K,2017MNRAS.470.4112G}, corresponding to periods of years.

In each of the observed and simulated light curves we identified the six most powerful GLS peaks.
In Fig.~\ref{fig:betadistns} we show the distribution of the powers of these  six peaks from our simulations with $\beta = -1.5,\;\beta = -1.9$, and $\beta = -2.5 $, compared to the observed powers in the corresponding peaks. The distribution for the case $\beta = -1.5$ matches the observed distribution most closely so we adopt that as our  fiducial case.

We next searched for independent harmonics in which the period ratios were 4/3, 3/2, 2/1, 3/1 and 4/1, where we counted as harmonics any match  within 10\% of one of the above ratios. or its reciprocal. The results are shown in Table \ref{tab:harmonics}.  We see that there are, on average, $4.5\pm0.23$ harmonics per observed light curve, and  $3.8\pm0.02$ harmonics per simulated light curve.  The ratio of the number of  harmonics per observed light curve to the number of harmonics per simulated light curve is therefore $1.17\pm 0.06$ which hints at a difference that is significant at just below the 3$\sigma$ level, where we are using $\sqrt{N}$ for our $\sigma$ estimates.  
It should be borne in mind that the simulations assume  a strict power law dependence whereas the actual shapes of the observed PSDs are more complex.  Thus there is some uncertainty that is not included in the uncertainty in the number of simulated harmonics in Table \ref{tab:harmonics}.  Nevertheless, especially given the hint of a harmonic that we found by hand in the analysis of the light curve of PKS J1309+1154, this slight difference should not be totally ignored.  

Our results with the other  assumed slopes of the PSD show that  power law slopes of $\beta = -1.9$ and $\beta = -2.5 $ return almost identical results to those assuming $\beta = -1.5$. Thus the occurrence of harmonics is independent of the power law slope over this range of $\beta$.

\begin{figure}[htp!]
    \centering
    \includegraphics[width=\linewidth]{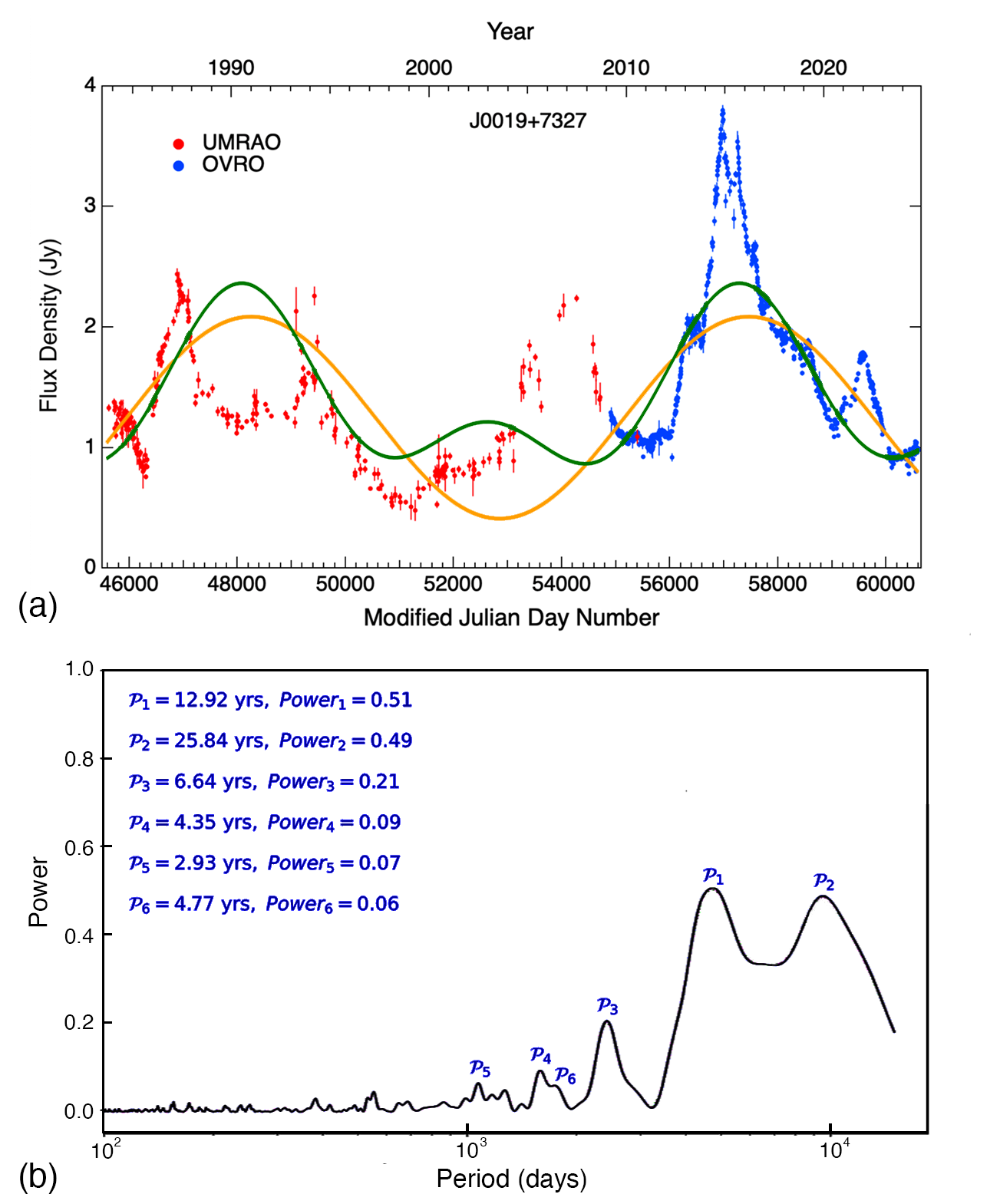}
    \caption{Light curve and GLS spectrum of J0019+7327. (a) The UMRAO+OVRO light curve: red points - UMRAO data, blue points - OVRO data. The orange curve shows the original fit of a sinusoid  to the data. The green curve shows the fit when a second sinusoidal component of half the period  is added (see text). (b) GLS spectrum  of the UMRAO+OVRO light curve shown in (a). The two brightest peaks have almost equal powers of $\mathcal{P}$=$\sim 50\%$ and their periods differ by a factor 2. }
     \label{fig:j0019lcandgls}
\end{figure}

Our conclusion from the above analysis is that the close agreement between the observed and the simulated number of harmonics per source is proof that the steep slope of the PSD produces random flares that are responsible for the vast majority of periodicities and  harmonics seen in  blazar GLS spectra.  

\begin{figure}[htp!]
    \centering
    \includegraphics[width=\linewidth]{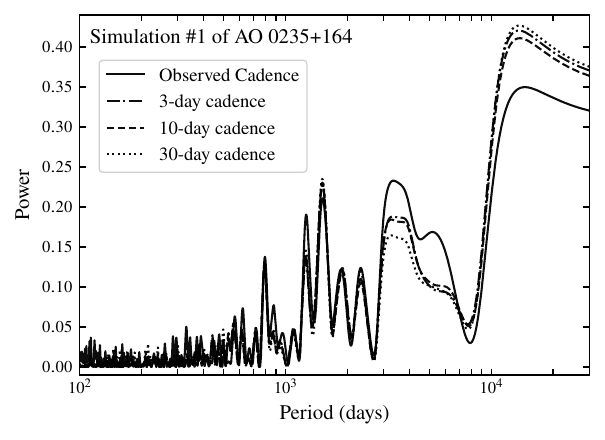}
    \caption{GLS spectra of simulations of the light curve of AO 0235+164. Since this is a simulation, we can sample it at any cadence of choice. The solid curve shows the data as actually sampled over 46 years in the UMRAO+OVRO program.  The other cadences are shown as indicated.  This makes it clear that the cadence has little effect on the peak locations in the GLS spectrum, provided that the light curve is well-sampled for  the periods under study.}
     \label{fig:ao0235}
\end{figure}

As an illustrative example, we carried out  the same least squares sine subtraction analysis of the light curve of J0019+7327 (see Fig.$\,$\ref{fig:j0019lcandgls}) that we used in the case of PKS J1309+1154 described in Section~\ref{sec:qpos}. We have chosen J0019+7327 because its GLS spectrum is particularly interesting, since its two strongest peaks have power $\mathcal{P}\approx 0.5$ with periods that differ by a factor 2.    We first fitted a simple sine-wave model and obtained a period of $25.2\pm0.2$ yr.  We then added a $P/2$ component and fitted both components simultaneously, allowing the period to vary, but maintaining the period ratio of 2. This yielded the period    $25.7\pm0.2$ yr.  The light curve and corresponding fits are shown in Fig.~\ref{fig:j0019lcandgls}(a).
It is clear  that the data do not strongly support the existence of these two sinusoidal periods.  This is borne out by the fact that when the first sinusoid (orange curve in Fig.\ref{fig:j0019lcandgls}(a)) is subtracted, the rms of the residual is reduced only by 30\%, and this is only reduced by a further 5\% when the two sinusoids (green curve in Fig.~\ref{fig:j0019lcandgls}(a)) are subtracted. For comparison, when the fitted sinusoid of PKS 2131--021 (Papers 1 and 2) is subtracted, the rms decreases by a factor 3. In PKS J1309+1154, when the linear trend and fitted sinusoid shown  in Fig.~\ref{plt:lightcurve1}(a) are subtracted the rms decreases by a factor 2.9, and when the harmonic is subtracted the rms scatter decreases by a further 24\%.

Finally, we tested whether the harmonics could be significantly affected by  the cadence of the observations, i.e., the average number of observations per unit time. In Fig.~\ref{fig:ao0235} we show the result of such a test on a simulated light curve of AO 0235+164. The cadence is clearly not responsible for producing the periodicities and harmonics we are seeing in our blazar light curves.

It is clear, therefore, that in the vast majority of cases, it is the random nature of blazar flares that leads to the multiple peaks in GLS spectra, frequently with apparent harmonic relationships, and therefore that neither of these is due to the dynamics of the SMBHs and their associated accretion disks.

\section{The Incidence of SMBHB Candidates Among Blazars}\label{incidence}

In this section we discuss two aspects of the numbers of SMBHB candidates among blazars.  We first compare the numbers of SMBHB candidate in the 16-year OVRO light curves with those in the 46-yr UMRAO+OVRO light curves, and we then consider the fractions of SMBHB candidates that our results imply.

\subsection{The Predicted Number of Strong SMBHB Candidates in the UMRAO+OVRO Light Curves of 83 Blazars}\label{sec:numbers}

The number of SMBHBs detected in a light curve monitoring program depends strongly on the duration $\tau$ of the observations. One can only detect SMBHBs having periods, $P$, less than a maximum, $P_\textrm{max}$, that is set by $\tau$. We adopt $P_\textrm{max}=0.3 \tau$, i.e., SMBHB candidates should be observed for  at least 3.5 cycles in order to be classified as strong candidates.

\citet{1980Natur.287..307B} first described the evolutionary stages of an SMBHB as the orbit shrinks.  Once the binary becomes sufficiently tightly bound,  the orbit is circularized, and gravitational radiation dominates the energy loss timescale. Thereafter gravitational radiation shrinks the orbit on a timescale $t_\textrm{GR}\propto r^4$, where $r$ is the separation of the black holes in the binary.
  
Let the number of SMBHBs, $N$, with separation less than $r$ be  $N(<r)$. Clearly  $N(<r) \propto r^4$ .  By Kepler's law  $r \propto P^{2/3}$. Thus, for any given combined SMBHB mass, $N(< P)\propto r^4 \propto P^{8/3}$. Therefore, in observations of duration $\tau_1$, \textit{vs.} observations of duration $\tau_2$,  the ratio of the numbers of sources in which we might expect to detect strong sinusoidal fluctuations,
$N_2/N_1 = [\tau_2/\tau_1]^{(8/3)}$.  

We note that this analysis does not account for non-gravitational torques that are potentially expected in accreting SMBHBs \citep{2023ARA&A..61..517L}, in particular systems with the moderately high Eddington ratios expected for BL Lac objects \citep[e.g.,][]{2007AJ....133.2187D}. Such torques may either arrest or hasten the binary orbital decay. 

Writing $N_2/N_1 = [\tau_2/\tau_1]^{\alpha}$, such torques can lead to $\alpha<8/3$, but they can also lead to $\alpha>8/3$. If dynamical forces accelerate systems to merge rapidly within a certain radius, then there should be a dearth of observable sources within that small radius, leading to $\alpha > 8/3$, but this could cause a build up of sources around some small radius as well, leading to  $\alpha<8/3$. It depends on the scenario,  but it probably will not deviate much from $\alpha = 8/3$ for SMBHBs with periods from  1-20 yr, since mergers can happen in $< 1$ Gyr for these sources.

Given that we have detected two SMBHB candidates out of 1830 in the OVRO sample, in the combined UMRAO+OVRO sample we expect to see strong sinusoidal variations in  1 out of $915 \times (16/46)^{8/3} = 55$ of the sources, i.e., we expect to see $ 83/55 \sim 1.5$ sources showing high amplitude sinusoidal variations over a large fraction of their 46-year light curves. One such source has already been found in this sample (PKS 2131--021), so there is an expectation of $\sim$0.43 that we will find a second such case in the combined sample, which is  a reasonably high probability. Thus, although based on small numbers, which is all we have to go on in this embryonic field at this very early stage,  there is a reasonable probability of detecting a second strong SMBHB candidate in the combined UMRAO+OVRO sample. 
 
 It was partly this consideration that motivated us to undertake the study of the combined UMRAO+OVRO sample -- we thought that the chances of finding a second SMBHB candidate were good, which, as this paper shows, has proven to be correct.

\subsection{Fractions of SMBHB Candidates in Blazar samples}\label{sec:fractions}

Paper 2 presented arguments suggesting that at least 1 in 100 blazars is an SMBHB candidate with orbital period in the range of months to 8 yr, where this upper limit is taken to be half the 16-yr duration of the light curves. In combining the UMRAO and OVRO light curves this upper limit on the period has been extended to 23 yr. While the combined UMRAO+OVRO sample is not a statistically well-defined sample, it is a sample of the brightest and most rapidly flaring blazars observable from the UMRAO and OVRO, and therefore of bright rapidly flaring blazars  north of declination $-20^\circ$. While PKS J1309+1154 requires confirmation as an SMBHB candidate, in our view the combination of its power in the GLS spectrum and the hint of a harmonic in its light curve, and the consistency of the ALMA data showing almost simultaneous but slightly leading sinusoidal variations to those at 15 GHz,  make it a very likely SMBHB candidate, so we will treat it as such.  In this case there are two SMBHB candidates  (PKS J1309+1154 and PKS 2131--021) in our sample of 83 blazars, i.e. a fraction of 2.4\%. It is useful to give a confidence interval for this fraction, which, for small numbers can be estimated in a number of ways.  The Wilson interval is $2.4_{-1.2}^{+2.3}\%$, whereas assuming Poisson statistics and applying Bayes' Theorem we find the interval to be $2.4_{-0.8}^{+3.2}\%$, which we adopt for this paper.

\section{Conclusion}\label{sec:conclusion}

The combined UMRAO+OVRO data set comprising continuous monitoring of 83 blazars  at 14.5/15 GHz wavelengths for $\sim$46--50 years is unprecedented in cadence and duration. It enables searches for periodicities of up to $\sim 20$ yr, and comparisons in independent long time intervals, as has been carried out here.   In the analysis of the variablity spectra of these 83 blazars, an in depth search for periodicities was carried out using the GLS, WWZ and SWF approaches. These revealed many apparent periodicities and harmonics.  In order to test the significance of these periodicities and their harmonics, we generated simulated light curves with similar power law PSD slopes to those observed in our sample.  These were then analyzed using the GLS method. \color{black} It was found that all but 6, 5, and 10 of the the GLS spectra of the simulated light curves in the $\beta\,=\,-1.5,\,-1.9,\;$  and $-2.5$ cases, respectively, showed multiple periodicities and harmonics.  In all three cases \color{black} we found four harmonics per source having ratios within 10\% of 4/3, 3/2, 2/1, 3/1, 4/1, or their reciprocals.

\textit{The fact that multiple harmonics are seen in \color{black} $\sim99.9\%$ of the \color{black} 24,900 simulations demonstrates clearly that random flares in blazar radio light curves  generate artificial periodicities and harmonics in abundance in GLS spectra.} 

Thus the only case that we are aware of where this approach can be used to study the dynamics of the central engines in blazars is that where the GLS spectrum is dominated by a single GLS peak of high power.  We recommend that all such studies use a power threshold  of $\mathcal{P}$=0.75 and a period allowing at least 2.5 cycles over the light curve window.

This study has discovered one new SMBHB candidate: PKS J1309+1154, with a radio light curve period of 17.9 yr, but since this does not reach our $3\sigma$ threshold in the GLS spectrum,  and since it has, in addition,  only been observed for 2.5 cycles, we do not yet consider it to be a strong SMBHB candidate. Since the period is 17.9 years it will take some time to observe another cycle and confirm it as a strong candidate, but this time may be halved if the harmonic is real.  In addition, optical observations might reveal evidence of shifting spectral lines that could confirm it as a strong candidate on a shorter timescale. 

Apart from the fact that its X-ray spectrum is significantly harder, (i) the  similarity of  the morphology of PKS J1309+1154 to that of PKS 2131--021, with  sinusoidal variations dominated by a bright core plus a close, stationary component that varies in concert with the core; (ii) the hint of a harmonic in PKS J1309+1154; and (iii) the small phase shift of the sinusoidal variations as a function of frequency, all support the interpretation of PKS J1309+1154 as an SMBHB candidate. It is clear, therefore, that PKS J1309+1154 is  well worth following up across the electromagnetic spectrum.  

Although the UMRAO+OVRO sample is not a carefully statistically selected sample, it is a sample of the brightest and most rapidly flaring blazars. Thus, if PKS J1309+1154 is confirmed as a strong SMBHB candidate, then in this cohort there would be 2/83 SMBHB candidates, implying a fraction of SMBHB candidates with orbital periods up to 20 yr amongst bright rapidly flaring blazars of $(2.4_{-0.8}^{+3.2})\%$.

\begin{acknowledgments}
This work is supported by NSF grants AST2407603 and AST2407604. We thank the California Institute of Technology and the Max Planck Institute for Radio Astronomy for supporting the  OVRO 40\,m program under extremely difficult circumstances over ten years (2014-2024) in the absence of agency funding for operation of the telescope. Without this private support this program would have ended in 2016.  We also thank all the volunteers who have enabled this work to be carried out.
Prior to~2016, the OVRO program was supported by NASA grants \hbox{NNG06GG1G}, \hbox{NNX08AW31G}, \hbox{NNX11A043G}, and \hbox{NNX13AQ89G} from~2006 to~2016 and NSF grants AST-0808050 and AST-1109911 from~2008 to~2014. The UMRAO program received support from NSF grants AST-8021250, AST-8301234, AST-8501093, AST-8815678, AST-9120224, AST-9421979, AST-9617032, AST-9900723, AST-0307629, AST-0607523, and earlier NSF awards, and from NASA grants NNX09AU16G, NNX10AP16G, NNX11AO13G, and NNX13AP18G. Additional funding for the operation of UMRAO was provided by the University of Michigan. 
W.{}M.\ acknowledges support from ANID projects Basal FB210003 and FONDECYT 11190853. A.S.  and R.B acknowledge support by a grant from the Simons Foundation (00001470,RB,AS). 
Y.{}D. and F.{}A.{}H.  acknowledge support through NASA under contract No. NNG08FD60C.
R.{}R.\ and B.{}M.\ and P.{}D.\ acknowledge support from ANID Basal AFB-170002, from Núcleo Milenio TITANs (NCN2023\_002), CATA BASAL FB210003 and UdeC-VRID 2025001479INV.
T.{}H.\ acknowledges support from the Academy of Finland projects 317383, 320085, 345899, and 362571 and from the European Union ERC-2024-COG - PARTICLES - 101169986.
I.L was funded by the European Union ERC-2022-STG - BOOTES - 101076343. Views and opinions expressed are however those of the author(s) only and do not necessarily reflect those of the European Union or the European Research Council Executive Agency. Neither the European Union nor the granting authority can be held responsible for them.

\color{black}
This paper depended on a very large amount of VLBI data, almost all of which was taken with the Very Long Baseline Array. The National Radio Astronomy Observatory is a facility of the National Science Foundation operated under cooperative agreement by Associated Universities, Inc. This paper makes use of the following ALMA data: ADS/JAO.ALMA\#2011.0.00001.CAL. ALMA is a partnership of ESO (representing its member states), NSF (USA) and NINS (Japan), together with NRC (Canada), NSTC and ASIAA (Taiwan), and KASI (Republic of Korea), in cooperation with the Republic of Chile. The Joint ALMA Observatory is operated by ESO, AUI/NRAO and NAOJ. This research has made use of the NASA/IPAC Extragalactic Database (NED), which is funded by the National Aeronautics and Space Administration and operated by the California Institute of Technology.
\color{black}
\end{acknowledgments}

\facilities{OVRO:40m, UMRAO, NRAO:VLBA, ALMA, ZTF, WISE}

    \clearpage
    \appendix
\twocolumngrid
\section{Sine Wave Fitting}\label{sec:sine}

We fitted a sine-wave model to the light curve of each of the 83 blazars using a methodology similar to that employed in Papers~1, 2, 3, and 4. In this model, the flux density is expressed as
\begin{equation}
S(t) = A \sin\left(\frac{2\pi (t - t_0)}{P} - \phi_0\right) + S_0,
\end{equation}
where $P$ is the sine-wave period, $A$ is its amplitude, $\phi_0$ is its phase at time $t=t_0$ (which is taken to be the mid-point of the observations for a given object), and $S_0$ is the mean flux density. The best-fit parameters were found by maximizing the following likelihood function:
\begin{equation}
\ln\mathcal{L} = -\frac{1}{2}\sum_i\left[\frac{(S_i - S(t_i))^2}{\sigma_i^2+\sigma_0^2} + \ln\left(\sigma_i^2+\sigma_0^2\right)\right],
\end{equation}
where $\sigma_0$ is a parameter that accounts for additional scatter in the light curves that is not captured by the original error bars (such as the intrinsic variability of the blazar). The posterior distributions for all model parameters were obtained using the Markov Chain Monte Carlo (MCMC) sampler by \citet{Foreman2013}. Note that this noise model is white noise, which does not include the effects of correlated noise, which can increase the uncertainties by a factor of up to 7. This point is discussed in detail in Appendix B of Paper 2.

One significant difference compared to Papers~1, 2, 3, and 4, is the choice of the starting points for the MCMC. For individual objects, it was possible to select the distributions of starting points by hand. However, for the entire sample of 83 blazars observed by both UMRAO and OVRO, this procedure had to be fully automated.

We used the following procedure. First, we kept the sine-wave period fixed and found the best-fit values of the parameters $(A, \phi_0,  S_0, \sigma_0)$ by maximizing the likelihood function using the MCMC. We recorded the maximum value of $\ln\mathcal{L}_{\rm max}$ and repeated the whole procedure for a range of periods (frequencies). The starting values of $A$ and $S_0$ were drawn from normal distributions with means $(\tilde{A}, \tilde{S}_0)$ and standard deviations $(0.1\tilde{A}, \sigma(\tilde{S}_0))$. Here, $2\tilde{A}$ represents the difference between the 95th and 5th percentiles of the flux density, $\tilde{S}_0$ is the mean flux density, and $\sigma(\tilde{S}_0)$ is its standard deviation. The starting values of $\phi_0$ and $\sigma_0$ were drawn from uniform distributions over the ranges $[0,2\pi]$ and $[0,\sigma(\tilde{S}_0)]$, respectively.  

Next, we calculated the best-fit parameters and the maximum likelihood for a range of frequencies, from $f_{\rm min} = 2 / \Delta T$ to $f_{\rm max} = 1\,\mathrm{yr}^{-1}$, using a step size of $\Delta f = 0.05 f_{\rm min}$, where $\Delta T$ is the time span of observations. We then selected the period (frequency) with the highest likelihood value and repeated the MCMC, this time allowing the period to vary. While doing the fits, we required the period to be shorter than $\Delta T$. We performed sine-wave fits for three data sets: combined UMRAO and OVRO, UMRAO alone, and OVRO alone. The results are reported in Table~\ref{tab:sine}.

\clearpage

\newpage

\startlongtable
\begin{deluxetable*}{lccc@{\hskip 8mm}lccc}
\tablecaption{Best-fit periods from sine-wave fits\label{tab:sine}}
\tablehead{
\colhead{Blazar} & \colhead{$P$ (days)} & \colhead{$P$ (days)} & \colhead{$P$ (days)} &
\colhead{Blazar} & \colhead{$P$ (days)} & \colhead{$P$ (days)} & \colhead{$P$ (days)} \\
\colhead{} & \colhead{(UMRAO)} & \colhead{(OVRO)} & \colhead{(combined)} &
\colhead{} & \colhead{(UMRAO)} & \colhead{(OVRO)} & \colhead{(combined)}
}
\tabletypesize{\footnotesize}
\startdata
J0010+1058 & $1868 \pm 8$ & $6920^{+400}_{-316}$ & $1845 \pm 5$ & J1217+3007 & $2518^{+65}_{-56}$ & $4843 \pm 86$ & $4401 \pm 47$ \\
J0019+7327 & $7064^{+143}_{-136}$ & $4666^{+61}_{-57}$ & $4616^{+46}_{-48}$ & J1221+2813 & $16720^{+34}_{-74}$ & $5891^{+199}_{-174}$ & $6309 \pm 66$ \\
J0050--0929 & $6294 \pm 244$ & $16264^{+156}_{-323}$ & $9076 \pm 225$ & C1224+2122 & $10602^{+478}_{-769}$ & $6213^{+110}_{-102}$ & $5922 \pm 46$ \\
0059+581 & $1632 \pm 24$ & $2697 \pm 29$ & $1699 \pm 8$ & J1229+0203 & $2987 \pm 13$ & $18192^{+169}_{-344}$ & $3065 \pm 14$ \\
J0108+0135 & $6686 \pm 102$ & $12039^{+2846}_{-1765}$ & $8694 \pm 116$ & J1256--0547 & $12788^{+254}_{-238}$ & $3118^{+30}_{-28}$ & $5220 \pm 64$ \\
J0111+3906 & $12075^{+1772}_{-1648}$ & $14811^{+116}_{-242}$ & $7702^{+130}_{-123}$ & J1305--1033 & $9956^{+905}_{-1127}$ & $2567 \pm 36$ & $2516 \pm 33$ \\
J0112+2244 & $8995 \pm 335$ & $15455^{+726}_{-1173}$ & $9275 \pm 104$ & J1309+1154 & $6391^{+115}_{-107}$ & $6201 \pm 48$ & $6509 \pm 32$ \\
J0136+4751 & $12655 \pm 198$ & $1478^{+14}_{-12}$ & $13148^{+210}_{-198}$ & J1310+3220 & $3362 \pm 26$ & $4414 \pm 21$ & $3559 \pm 18$ \\
J0204+1514 & $16077^{+178}_{-370}$ & $16195^{+91}_{-194}$ & $3908 \pm 27$ & J1337--1257 & $2647 \pm 21$ & $2039 \pm 14$ & $2521 \pm 11$ \\
J0217+7349 & $4865 \pm 80$ & $2676 \pm 17$ & $2786 \pm 18$ & J1415+1320 & $3206 \pm 73$ & $3670 \pm 60$ & $3262 \pm 16$ \\
J0217+0144 & $3007 \pm 27$ & $3971 \pm 40$ & $4082 \pm 27$ & J1419+5423 & $16512^{+73}_{-154}$ & $4159 \pm 47$ & $4384 \pm 22$ \\
0224+671 & $14456^{+810}_{-1237}$ & $1829 \pm 14$ & $1914 \pm 6$ & PKS1510--089 & $9859 \pm 319$ & $1687^{+21}_{-20}$ & $9780 \pm 197$ \\
J0237+2848 & $7084 \pm 129$ & $9934^{+1031}_{-734}$ & $4942 \pm 38$ & J1540+1447 & $8264^{+307}_{-263}$ & $2997^{+40}_{-38}$ & $8408 \pm 67$ \\
J0238+1636 & $1988 \pm 12$ & $2501^{+36}_{-34}$ & $2062 \pm 7$ & J1555+1111 & $14314^{+567}_{-756}$ & $10789^{+759}_{-605}$ & $6550^{+85}_{-79}$ \\
J0259+0747 & $14909^{+1104}_{-1637}$ & $11530 \pm 3357$ & $13581^{+306}_{-290}$ & J1613+3412 & $13177^{+374}_{-341}$ & $5348 \pm 44$ & $7180 \pm 36$ \\
0300+471 & $16416^{+378}_{-528}$ & $15379^{+438}_{-744}$ & $6198 \pm 37$ & J1635+3808 & $9853 \pm 146$ & $967 \pm 6$ & $982 \pm 3$ \\
J0309+1029 & $3934^{+56}_{-53}$ & $5508^{+110}_{-98}$ & $6860 \pm 37$ & J1642+6856 & $15952^{+734}_{-1232}$ & $16632^{+246}_{-500}$ & $6377 \pm 68$ \\
J0319+4130 & $17186 \pm 100$ & $18347^{+109}_{-230}$ & $18464^{+21}_{-44}$ & J1642+3948 & $5966 \pm 39$ & $17965^{+335}_{-655}$ & $5608 \pm 48$ \\
0333+321 & $15686^{+460}_{-685}$ & $15456^{+644}_{-1091}$ & $12781 \pm 156$ & J1653+3945 & $12622^{+531}_{-437}$ & $2503 \pm 33$ & $2305 \pm 11$ \\
J0339--0146 & $4277 \pm 74$ & $5990 \pm 73$ & $6335 \pm 36$ & J1719+1745 & $2763 \pm 26$ & $7531^{+332}_{-277}$ & $7187 \pm 75$ \\
0415+379 & $4186^{+49}_{-53}$ & $1539 \pm 12$ & $3784 \pm 30$ & J1733--1304 & $10261^{+416}_{-355}$ & $2866 \pm 39$ & $3767 \pm 32$ \\
J0423--0120 & $2979 \pm 22$ & $4537 \pm 35$ & $7716 \pm 85$ & J1740+5211 & $9731^{+517}_{-444}$ & $1733 \pm 13$ & $1723 \pm 5$ \\
C0424+0036 & $7908 \pm 168$ & $16255^{+163}_{-338}$ & $9094 \pm 71$ & J1743--0350 & $9989^{+532}_{-431}$ & $1650 \pm 8$ & $2745 \pm 13$ \\
J0433+0521 & $3985 \pm 32$ & $5706^{+98}_{-93}$ & $3932 \pm 21$ & J1748+7005 & $16108^{+125}_{-263}$ & $13770^{+1676}_{-1841}$ & $6252 \pm 34$ \\
J0501--0159 & $13226^{+114}_{-235}$ & $11151^{+996}_{-802}$ & $6074 \pm 87$ & J1751+0939 & $16315^{+321}_{-556}$ & $14755^{+1404}_{-1940}$ & $848 \pm 3$ \\
J0530+1331 & $3761 \pm 40$ & $15895^{+150}_{-311}$ & $3953^{+40}_{-38}$ & J1800+7828 & $13780^{+1065}_{-910}$ & $1944^{+48}_{-43}$ & $9821 \pm 198$ \\
J0607--0834 & $12873^{+678}_{-572}$ & $4794 \pm 56$ & $9242 \pm 72$ & J1806+6949 & $14235^{+946}_{-831}$ & $6612^{+139}_{-129}$ & $6097 \pm 69$ \\
J0609--1542 & $16442^{+131}_{-271}$ & $2198 \pm 23$ & $4184 \pm 33$ & J1824+5651 & $6840^{+201}_{-176}$ & $3355 \pm 25$ & $3022 \pm 17$ \\
J0721+7120 & $13095^{+646}_{-541}$ & $12042^{+2175}_{-1680}$ & $14777 \pm 224$ & J1927+7358 & $4855^{+91}_{-84}$ & $5133^{+130}_{-115}$ & $7924 \pm 67$ \\
PKS0727--115 & $6534 \pm 35$ & $17880^{+238}_{-491}$ & $7356 \pm 54$ & J2005+7752 & $11496^{+436}_{-404}$ & $2751 \pm 26$ & $6749 \pm 81$ \\
J0738+1742 & $11531^{+509}_{-444}$ & $3265 \pm 22$ & $4252 \pm 32$ & 2005+403 & $17638^{+310}_{-486}$ & $4263 \pm 44$ & $8768 \pm 45$ \\
J0739+0137 & $6096 \pm 119$ & $5298^{+119}_{-110}$ & $6536 \pm 68$ & J2022+6136 & $7430^{+284}_{-256}$ & $10956^{+1554}_{-1121}$ & $6262 \pm 95$ \\
J0757+0956 & $3637^{+62}_{-55}$ & $7801^{+940}_{-629}$ & $8138 \pm 74$ & J2123+0535 & $2403 \pm 15$ & $5961^{+125}_{-117}$ & $7184 \pm 89$ \\
J0808+4950 & $8061^{+1980}_{-1014}$ & $11426^{+2707}_{-1929}$ & $4260 \pm 38$ & J2134--0153 & $10735^{+1291}_{-639}$ & $1787 \pm 5$ & $1723 \pm 3$ \\
J0818+4222 & $14068^{+690}_{-580}$ & $3800 \pm 24$ & $4632 \pm 24$ & J2158--1501 & $2013 \pm 33$ & $16044^{+174}_{-365}$ & $16246^{+24}_{-52}$ \\
J0831+0429 & $12916^{+980}_{-748}$ & $14549^{+1342}_{-1796}$ & $6501 \pm 76$ & C2225--0457 & $3097 \pm 16$ & $16320^{+74}_{-158}$ & $9066 \pm 163$ \\
J0854+2006 & $10276 \pm 100$ & $2124 \pm 25$ & $12569^{+177}_{-167}$ & 2230+114 & $12002^{+237}_{-220}$ & $15389^{+746}_{-1354}$ & $1464 \pm 5$ \\
J0958+6533 & $8941^{+516}_{-422}$ & $13226^{+275}_{-554}$ & $6741 \pm 103$ & J2236+2828 & $1087^{+3}_{-3}$ & $6578^{+208}_{-181}$ & $6504^{+161}_{-147}$ \\
J1058+0133 & $6152 \pm 86$ & $6490^{+236}_{-205}$ & $5395 \pm 62$ & J2253+1608 & $2112 \pm 7$ & $2768 \pm 21$ & $2735 \pm 10$ \\
J1104+3812 & $16211^{+292}_{-576}$ & $4908 \pm 73$ & $5116^{+65}_{-68}$ & 2254+074 & $8179^{+649}_{-467}$ & $3066^{+51}_{-48}$ & $3466^{+39}_{-41}$ \\
J1150+2417 & $11523^{+2290}_{-1505}$ & $13457^{+1028}_{-837}$ & $9333 \pm 94$ & J2348--1631 & $1384^{+20}_{-16}$ & $1283 \pm 9$ & $1306 \pm 6$ \\
J1159+2914 & $1237 \pm 6$ & $2396 \pm 23$ & $2760 \pm 13$ & \nodata & \nodata & \nodata & \nodata \\
\enddata
\end{deluxetable*}

\bibliography{references}{}

\begin{thebibliography}{}
\expandafter\ifx\csname natexlab\endcsname\relax\def\natexlab#1{#1}\fi
\providecommand{\url}[1]{\href{#1}{#1}}
\providecommand{\dodoi}[1]{doi:~\href{http://doi.org/#1}{\nolinkurl{#1}}}
\providecommand{\doeprint}[1]{\href{http://ascl.net/#1}{\nolinkurl{http://ascl.net/#1}}}
\providecommand{\doarXiv}[1]{\href{https://arxiv.org/abs/#1}{\nolinkurl{https://arxiv.org/abs/#1}}}

\bibitem[{N. {Agarwal} {et~al.}(2025){Agarwal}, {Agazie}, {Anumarlapudi}, {Archibald}, {Arzoumanian}, {Baier}, {Baker}, {Becsy}, {Blecha}, {Brazier}, {Brook}, {Burke-Spolaor}, {Burnette}, {Case}, {Casey-Clyde}, {Chang}, {Charisi}, {Chatterjee}, {Cohen}, {Coppi}, {Cordes}, {Cornish}, {Crawford}, {Cromartie}, {Crowter}, {DeCesar}, {Demorest}, {Deng}, {Dey}, {Dolch}, {D'Orazio}, {Eisenberg}, {Ferrara}, {Fiore}, {Fonseca}, {Freedman}, {Gardiner}, {Garver-Daniels}, {Gentile}, {Gersbach}, {Glaser}, {Graham}, {Good}, {Gultekin}, {Harris}, {Hazboun}, {Hutchison}, {Jennings}, {Johnson}, {Jones}, {Kaplan}, {Kelley}, {Kerr}, {Key}, {Laal}, {Lam}, {Lamb}, {Larsen}, {Lazio}, {Lewandowska}, {Liu}, {Lorimer}, {Luo}, {Lynch}, {Ma}, {Madison}, {Matt}, {McEwen}, {McKee}, {McLaughlin}, {McMann}, {Meyers}, {Meyers}, {Mingarelli}, {Mitridate}, {Natarajan}, {Ng}, {Nice}, {Ocker}, {Olum}, {Pennucci}, {Perera}, {Petrov}, {Pol}, {Radovan}, {Ransom}, {Ray}, {Romano}, {Runnoe}, {Saffer}, {Sardesai}, {Schmiedekamp}, {Schmiedekamp},
  {Schmitz}, {Semenzato}, {Shapiro-Albert}, {Shivakumar}, {Siemens}, {Simon}, {Sosa Fiscella}, {Stairs}, {Stinebring}, {Stovall}, {Susobhanan}, {Swiggum}, {Taylor}, {Taylor}, {Thompson}, {Turner}, {Vallisneri}, {van Haasteren}, {Vigeland}, {Wahl}, {Willson}, {Wilson}, {Witt}, {Wright}, {Young}, \& {Zheng}}]{2025arXiv250816534A}
{Agarwal}, N., {Agazie}, G., {Anumarlapudi}, A., {et~al.} 2025, \bibinfo{title}{{The NANOGrav 15 yr Data Set: Targeted Searches for Supermassive Black Hole Binaries},} arXiv e-prints, arXiv:2508.16534, \dodoi{10.48550/arXiv.2508.16534}

\bibitem[{G. {Agazie} {et~al.}(2023){Agazie}, {Anumarlapudi}, {Archibald}, {Arzoumanian}, {Baker}, {B{\'e}csy}, {Blecha}, {Brazier}, {Brook}, {Burke-Spolaor}, {Burnette}, {Case}, {Charisi}, {Chatterjee}, {Chatziioannou}, {Cheeseboro}, {Chen}, {Cohen}, {Cordes}, {Cornish}, {Crawford}, {Cromartie}, {Crowter}, {Cutler}, {Decesar}, {Degan}, {Demorest}, {Deng}, {Dolch}, {Drachler}, {Ellis}, {Ferrara}, {Fiore}, {Fonseca}, {Freedman}, {Garver-Daniels}, {Gentile}, {Gersbach}, {Glaser}, {Good}, {G{\"u}ltekin}, {Hazboun}, {Hourihane}, {Islo}, {Jennings}, {Johnson}, {Jones}, {Kaiser}, {Kaplan}, {Kelley}, {Kerr}, {Key}, {Klein}, {Laal}, {Lam}, {Lamb}, {Lazio}, {Lewandowska}, {Littenberg}, {Liu}, {Lommen}, {Lorimer}, {Luo}, {Lynch}, {Ma}, {Madison}, {Mattson}, {McEwen}, {McKee}, {McLaughlin}, {McMann}, {Meyers}, {Meyers}, {Mingarelli}, {Mitridate}, {Natarajan}, {Ng}, {Nice}, {Ocker}, {Olum}, {Pennucci}, {Perera}, {Petrov}, {Pol}, {Radovan}, {Ransom}, {Ray}, {Romano}, {Sardesai}, {Schmiedekamp}, {Schmiedekamp}, {Schmitz},
  {Schult}, {Shapiro-Albert}, {Siemens}, {Simon}, {Siwek}, {Stairs}, {Stinebring}, {Stovall}, {Sun}, {Susobhanan}, {Swiggum}, {Taylor}, {Taylor}, {Turner}, {Unal}, {Vallisneri}, {van Haasteren}, {Vigeland}, {Wahl}, {Wang}, {Witt}, {Young}, \& {Nanograv Collaboration}}]{2023ApJ...951L...8A}
{Agazie}, G., {Anumarlapudi}, A., {Archibald}, A.~M., {et~al.} 2023, \bibinfo{title}{{The NANOGrav 15 yr Data Set: Evidence for a Gravitational-wave Background},} \apjl, 951, L8, \dodoi{10.3847/2041-8213/acdac6}

\bibitem[{F.~D. {Albareti} {et~al.}(2017){Albareti}, {Allende Prieto}, {Almeida}, {Anders}, {Anderson}, {Andrews}, {Arag{\'o}n-Salamanca}, {Argudo-Fern{\'a}ndez}, {Armengaud}, {Aubourg}, {Avila-Reese}, {Badenes}, {Bailey}, {Barbuy}, {Barger}, {Barrera-Ballesteros}, {Bartosz}, {Basu}, {Bates}, {Battaglia}, {Baumgarten}, {Baur}, {Bautista}, {Beers}, {Belfiore}, {Bershady}, {Bertran de Lis}, {Bird}, {Bizyaev}, {Blanc}, {Blanton}, {Blomqvist}, {Bolton}, {Borissova}, {Bovy}, {Brandt}, {Brinkmann}, {Brownstein}, {Bundy}, {Burtin}, {Busca}, {Camacho Chavez}, {Cano D{\'\i}az}, {Cappellari}, {Carrera}, {Chen}, {Cherinka}, {Cheung}, {Chiappini}, {Chojnowski}, {Chuang}, {Chung}, {Cirolini}, {Clerc}, {Cohen}, {Comerford}, {Comparat}, {Correa do Nascimento}, {Cousinou}, {Covey}, {Crane}, {Croft}, {Cunha}, {Darling}, {Davidson}, {Dawson}, {Da Costa}, {Da Silva Ilha}, {Deconto Machado}, {Delubac}, {De Lee}, {De la Macorra}, {De la Torre}, {Diamond-Stanic}, {Donor}, {Downes}, {Drory}, {Du}, {Du Mas des Bourboux}, {Dwelly},
  {Ebelke}, {Eigenbrot}, {Eisenstein}, {Elsworth}, {Emsellem}, {Eracleous}, {Escoffier}, {Evans}, {Falc{\'o}n-Barroso}, {Fan}, {Favole}, {Fernandez-Alvar}, {Fernandez-Trincado}, {Feuillet}, {Fleming}, {Font-Ribera}, {Freischlad}, {Frinchaboy}, {Fu}, {Gao}, {Garcia}, {Garcia-Dias}, {Garcia-Hern{\'a}ndez}, {Garcia P{\'e}rez}, {Gaulme}, {Ge}, {Geisler}, {Gillespie}, {Gil Marin}, {Girardi}, {Goddard}, {Gomez Maqueo Chew}, {Gonzalez-Perez}, {Grabowski}, {Green}, {Grier}, {Grier}, {Guo}, {Guy}, {Hagen}, {Hall}, {Harding}, {Harley}, {Hasselquist}, {Hawley}, {Hayes}, {Hearty}, {Hekker}, {Hernandez Toledo}, {Ho}, {Hogg}, {Holley-Bockelmann}, {Holtzman}, {Holzer}, {Hu}, {Huber}, {Hutchinson}, {Hwang}, {Ibarra-Medel}, {Ivans}, {Ivory}, {Jaehnig}, {Jensen}, {Johnson}, {Jones}, {Jullo}, {Kallinger}, {Kinemuchi}, {Kirkby}, {Klaene}, {Kneib}, {Kollmeier}, {Lacerna}, {Lane}, {Lang}, {Laurent}, {Law}, {Leauthaud}, {Le Goff}, {Li}, {Li}, {Li}, {Li}, {Liang}, {Liang}, {Lima}, {Lin}, {Lin}, {Lin}, {Liu}, {Long}, {Lucatello},
  {MacDonald}, {MacLeod}, {Mackereth}, {Mahadevan}, {Maia}, {Maiolino}, {Majewski}, {Malanushenko}, {Malanushenko}, {Mallmann}, {Manchado}, {Maraston}, {Marques-Chaves}, {Martinez Valpuesta}, {Masters}, {Mathur}, {McGreer}, {Merloni}, {Merrifield}, {M{\'e}sz{\'a}ros}, {Meza}, {Miglio}, {Minchev}, {Molaverdikhani}, {Montero-Dorta}, {Mosser}, {Muna}, \& {Myers}}]{2017ApJS..233...25A}
{Albareti}, F.~D., {Allende Prieto}, C., {Almeida}, A., {et~al.} 2017, \bibinfo{title}{{The 13th Data Release of the Sloan Digital Sky Survey: First Spectroscopic Data from the SDSS-IV Survey Mapping Nearby Galaxies at Apache Point Observatory},} \apjs, 233, 25, \dodoi{10.3847/1538-4365/aa8992}

\bibitem[{H.~D. {Aller} {et~al.}(1985){Aller}, {Aller}, {Latimer}, \& {Hodge}}]{1985ApJS...59..513A}
{Aller}, H.~D., {Aller}, M.~F., {Latimer}, G.~E., \& {Hodge}, P.~E. 1985, \bibinfo{title}{{Spectra and linear polarizations of extragalactic variable sources atcentimeter wavelengths.},} \apjs, 59, 513, \dodoi{10.1086/191083}

\bibitem[{M.~F. {Aller} {et~al.}(2014){Aller}, {Hughes}, {Aller}, {Latimer}, \& {Hovatta}}]{2014ApJ...791...53A}
{Aller}, M.~F., {Hughes}, P.~A., {Aller}, H.~D., {Latimer}, G.~E., \& {Hovatta}, T. 2014, \bibinfo{title}{{Constraining the Physical Conditions in the Jets of {\ensuremath{\gamma}}-Ray Flaring Blazars Using Centimeter-band Polarimetry and Radiative Transfer Simulations. I. Data and Models for 0420-014, OJ 287, and 1156+295},} \apj, 791, 53, \dodoi{10.1088/0004-637X/791/1/53}

\bibitem[{T. {An} {et~al.}(2013){An}, {Baan}, {Wang}, {Wang}, \& {Hong}}]{2013MNRAS.434.3487A}
{An}, T., {Baan}, W.~A., {Wang}, J.-Y., {Wang}, Y., \& {Hong}, X.-Y. 2013, \bibinfo{title}{{Periodic radio variabilities in NRAO 530: a jet-disc connection?},} \mnras, 434, 3487, \dodoi{10.1093/mnras/stt1265}

\bibitem[{K.~A. {Arnaud}(1996){Arnaud}}]{Arnaud1996}
{Arnaud}, K.~A. 1996, \bibinfo{title}{{XSPEC: The First Ten Years},} in ASP Conf. Ser., Vol. 101, Astronomical Data Analysis Software and Systems V, ed. G.~H. {Jacoby} \& J.~{Barnes} (San Francisco: ASP), 17

\bibitem[{T.~G. {Arshakian} {et~al.}(2024){Arshakian}, {Hambardzumyan}, {Pushkarev}, \& {Homan}}]{2024AandA...692A.127A}
{Arshakian}, T.~G., {Hambardzumyan}, L.~A., {Pushkarev}, A.~B., \& {Homan}, D.~C. 2024, \bibinfo{title}{{Studies of stationary features in jets: BL Lacertae: II. Trajectory reversals and superluminal speeds on sub-parsec scales},} \aap, 692, A127, \dodoi{10.1051/0004-6361/202451406}

\bibitem[{T.~G. {Arshakian} {et~al.}(2025){Arshakian}, {Hambardzumyan}, {Pushkarev}, {Homan}, \& {Karapetyan}}]{2025arXiv250612457A}
{Arshakian}, T.~G., {Hambardzumyan}, L.~A., {Pushkarev}, A.~B., {Homan}, D.~C., \& {Karapetyan}, E.~L. 2025, \bibinfo{title}{{Studies of stationary features in jets: 3C 279 quasar I. On-sky scattering and dynamics},} arXiv e-prints, arXiv:2506.12457, \dodoi{10.48550/arXiv.2506.12457}

\bibitem[{T.~G. {Arshakian} {et~al.}(2020){Arshakian}, {Pushkarev}, {Lister}, \& {Savolainen}}]{2020AandA...640A..62A}
{Arshakian}, T.~G., {Pushkarev}, A.~B., {Lister}, M.~L., \& {Savolainen}, T. 2020, \bibinfo{title}{{Studies of stationary features in jets: BL Lacertae. I. The dynamics and brightness asymmetry on sub-parsec scales},} \aap, 640, A62, \dodoi{10.1051/0004-6361/202037968}

\bibitem[{J.~W.~M. {Baars} {et~al.}(1977){Baars}, {Genzel}, {Pauliny-Toth}, \& {Witzel}}]{Baars1977}
{Baars}, J. W.~M., {Genzel}, R., {Pauliny-Toth}, I. I.~K., \& {Witzel}, A. 1977, \bibinfo{title}{{The absolute spectrum of Cas A: an accurate flux density scale and a set of secondary calibrators.},} \aap, 61, 99 , \dodoi{1977A&A....61...99B}

\bibitem[{R.~V. {Baluev}(2009){Baluev}}]{2009MNRAS.395.1541B}
{Baluev}, R.~V. 2009, \bibinfo{title}{{Detecting non-sinusoidal periodicities in observational data using multiharmonic periodograms},} \mnras, 395, 1541, \dodoi{10.1111/j.1365-2966.2009.14634.x}

\bibitem[{A. {Banerjee} {et~al.}(2023){Banerjee}, {Sharma}, {Mandal}, {Das}, {Bhatta}, \& {Bose}}]{2023MNRAS.523L..52B}
{Banerjee}, A., {Sharma}, A., {Mandal}, A., {et~al.} 2023, \bibinfo{title}{{Detection of periodicity in the gamma-ray light curve of the BL Lac 4FGL J2202.7+4216},} \mnras, 523, L52, \dodoi{10.1093/mnrasl/slad057}

\bibitem[{M.~C. {Begelman} {et~al.}(1980){Begelman}, {Blandford}, \& {Rees}}]{1980Natur.287..307B}
{Begelman}, M.~C., {Blandford}, R.~D., \& {Rees}, M.~J. 1980, \bibinfo{title}{{Massive black hole binaries in active galactic nuclei},} \nat, 287, 307, \dodoi{10.1038/287307a0}

\bibitem[{E.~C. {Bellm} {et~al.}(2019){Bellm}, {Kulkarni}, {Graham}, {Dekany}, {Smith}, {Riddle}, {Masci}, {Helou}, {Prince}, {Adams}, {Barbarino}, {Barlow}, {Bauer}, {Beck}, {Belicki}, {Biswas}, {Blagorodnova}, {Bodewits}, {Bolin}, {Brinnel}, {Brooke}, {Bue}, {Bulla}, {Burruss}, {Cenko}, {Chang}, {Connolly}, {Coughlin}, {Cromer}, {Cunningham}, {De}, {Delacroix}, {Desai}, {Duev}, {Eadie}, {Farnham}, {Feeney}, {Feindt}, {Flynn}, {Franckowiak}, {Frederick}, {Fremling}, {Gal-Yam}, {Gezari}, {Giomi}, {Goldstein}, {Golkhou}, {Goobar}, {Groom}, {Hacopians}, {Hale}, {Henning}, {Ho}, {Hover}, {Howell}, {Hung}, {Huppenkothen}, {Imel}, {Ip}, {Ivezi{\'c}}, {Jackson}, {Jones}, {Juric}, {Kasliwal}, {Kaspi}, {Kaye}, {Kelley}, {Kowalski}, {Kramer}, {Kupfer}, {Landry}, {Laher}, {Lee}, {Lin}, {Lin}, {Lunnan}, {Giomi}, {Mahabal}, {Mao}, {Miller}, {Monkewitz}, {Murphy}, {Ngeow}, {Nordin}, {Nugent}, {Ofek}, {Patterson}, {Penprase}, {Porter}, {Rauch}, {Rebbapragada}, {Reiley}, {Rigault}, {Rodriguez}, {van Roestel}, {Rusholme},
  {van Santen}, {Schulze}, {Shupe}, {Singer}, {Soumagnac}, {Stein}, {Surace}, {Sollerman}, {Szkody}, {Taddia}, {Terek}, {Van Sistine}, {van Velzen}, {Vestrand}, {Walters}, {Ward}, {Ye}, {Yu}, {Yan}, \& {Zolkower}}]{2019PASP..131a8002B}
{Bellm}, E.~C., {Kulkarni}, S.~R., {Graham}, M.~J., {et~al.} 2019, \bibinfo{title}{{The Zwicky Transient Facility: System Overview, Performance, and First Results},} \pasp, 131, 018002, \dodoi{10.1088/1538-3873/aaecbe}

\bibitem[{G. {Bhatta}(2017){Bhatta}}]{2017ApJ...847....7B}
{Bhatta}, G. 2017, \bibinfo{title}{{Radio and {\ensuremath{\gamma}}-Ray Variability in the BL Lac PKS 0219-164: Detection of Quasi-periodic Oscillations in the Radio Light Curve},} \apj, 847, 7, \dodoi{10.3847/1538-4357/aa86ed}

\bibitem[{R. {Blandford} {et~al.}(2019){Blandford}, {Meier}, \& {Readhead}}]{2019ARAandA..57..467B}
{Blandford}, R., {Meier}, D., \& {Readhead}, A. 2019, \bibinfo{title}{{Relativistic Jets from Active Galactic Nuclei},} \araa, 57, 467, \dodoi{10.1146/annurev-astro-081817-051948}

\bibitem[{G.~L. {Bretthorst}(2003){Bretthorst}}]{2003sca..book..309B}
{Bretthorst}, G.~L. 2003, \bibinfo{title}{{Frequency estimation and generalized Lomb-Scargle periodograms},} in Statistical Challenges in Astronomy, ed. E.~D. {Feigelson} \& G.~J. {Babu}, 309--329

\bibitem[{V.~S. {Bychkova} {et~al.}(2015){Bychkova}, {Vol'vach}, {Kardashev}, {Larionov}, {Vlasyuk}, {Spiridonova}, {Vol'vach}, {L{\"a}hteenm{\"a}ki}, {Tornikoski}, {Aller}, \& {Aller}}]{2015ARep...59..851B}
{Bychkova}, V.~S., {Vol'vach}, A.~E., {Kardashev}, N.~S., {et~al.} 2015, \bibinfo{title}{{Long-term monitoring of the blazars AO 0235+164 and S5 0716+714 in the optical and radio ranges},} Astronomy Reports, 59, 851, \dodoi{10.1134/S1063772915080016}

\bibitem[{M.~H. {Cohen} {et~al.}(2014){Cohen}, {Meier}, {Arshakian}, {Homan}, {Hovatta}, {Kovalev}, {Lister}, {Pushkarev}, {Richards}, \& {Savolainen}}]{2014ApJ...787..151C}
{Cohen}, M.~H., {Meier}, D.~L., {Arshakian}, T.~G., {et~al.} 2014, \bibinfo{title}{{Studies of the Jet in Bl Lacertae. I. Recollimation Shock and Moving Emission Features},} \apj, 787, 151, \dodoi{10.1088/0004-637X/787/2/151}

\bibitem[{M.~H. {Cohen} {et~al.}(2015){Cohen}, {Meier}, {Arshakian}, {Clausen-Brown}, {Homan}, {Hovatta}, {Kovalev}, {Lister}, {Pushkarev}, {Richards}, \& {Savolainen}}]{2015ApJ...803....3C}
{Cohen}, M.~H., {Meier}, D.~L., {Arshakian}, T.~G., {et~al.} 2015, \bibinfo{title}{{Studies of the Jet in Bl Lacertae. II. Superluminal Alfv{\'e}n Waves},} \apj, 803, 3, \dodoi{10.1088/0004-637X/803/1/3}

\bibitem[{H. {Dai} {et~al.}(2007){Dai}, {Xie}, {Zhou}, {Li}, {Chen}, \& {Ma}}]{2007AJ....133.2187D}
{Dai}, H., {Xie}, G.~Z., {Zhou}, S.~B., {et~al.} 2007, \bibinfo{title}{{Correlation between Eddington Ratios and Broad-Line Luminosities in Flat-Spectrum Radio Quasars, BL Lacertae Objects, and Radio Galaxies},} \aj, 133, 2187, \dodoi{10.1086/511769}

\bibitem[{P.~V. {de la Parra} {et~al.}(2025){de la Parra}, {Kiehlmann}, {Mr{\'o}z}, {Readhead}, {Synani}, {Begelman}, {Blandford}, {Ding}, {Harrison}, {Liodakis}, {Max-Moerbeck}, {Pavlidou}, {Reeves}, {Vallisneri}, {Aller}, {Graham}, {Hovatta}, {Lawrence}, {Lazio}, {Mahabal}, {Molina}, {O'Neill}, {Pearson}, {Ravi}, {Tassis}, \& {Zensus}}]{2025ApJ...987..191D}
{de la Parra}, P.~V., {Kiehlmann}, S., {Mr{\'o}z}, P., {et~al.} 2025, \bibinfo{title}{{PKS J0805-0111: A Second Owens Valley Radio Observatory Blazar Showing Highly Significant Sinusoidal Radio Variability{\textemdash}The Tip of the Iceberg},} \apj, 987, 191, \dodoi{10.3847/1538-4357/addc60}

\bibitem[{F.-T. {Dong} {et~al.}(2022){Dong}, {Gai}, {Tang}, {Wang}, \& {Yi}}]{2022RAA....22k5001D}
{Dong}, F.-T., {Gai}, N., {Tang}, Y., {Wang}, Y.-F., \& {Yi}, T.-F. 2022, \bibinfo{title}{{Evidence of Quasi-periodic Oscillation in the Optical Band of the Blazar 1ES 1959+650},} Research in Astronomy and Astrophysics, 22, 115001, \dodoi{10.1088/1674-4527/ac71fc}

\bibitem[{A.~J. {Drake} {et~al.}(2009){Drake}, {Djorgovski}, {Mahabal}, {Beshore}, {Larson}, {Graham}, {Williams}, {Christensen}, {Catelan}, {Boattini}, {Gibbs}, {Hill}, \& {Kowalski}}]{Drake2009}
{Drake}, A.~J., {Djorgovski}, S.~G., {Mahabal}, A., {et~al.} 2009, \bibinfo{title}{{First Results from the Catalina Real-Time Transient Survey},} \apj, 696, 870, \dodoi{10.1088/0004-637X/696/1/870}

\bibitem[{ {EPTA Collaboration}(2023){EPTA Collaboration}}]{epta:2023}
{EPTA Collaboration}. 2023, \bibinfo{title}{{The second data release from the European Pulsar Timing Array. III. Search for gravitational wave signals},} \aap, 678, A50, \dodoi{10.1051/0004-6361/202346844}

\bibitem[{ {Event Horizon Telescope Collaboration} {et~al.}(2019){Event Horizon Telescope Collaboration}, {Akiyama}, {Alberdi}, {Alef}, {Asada}, {Azulay}, {Baczko}, {Ball}, {Balokovi{\'c}}, {Barrett}, {Bintley}, {Blackburn}, {Boland}, {Bouman}, {Bower}, {Bremer}, {Brinkerink}, {Brissenden}, {Britzen}, {Broderick}, {Broguiere}, {Bronzwaer}, {Byun}, {Carlstrom}, {Chael}, {Chan}, {Chatterjee}, {Chatterjee}, {Chen}, {Chen}, {Cho}, {Christian}, {Conway}, {Cordes}, {Crew}, {Cui}, {Davelaar}, {De Laurentis}, {Deane}, {Dempsey}, {Desvignes}, {Dexter}, {Doeleman}, {Eatough}, {Falcke}, {Fish}, {Fomalont}, {Fraga-Encinas}, {Freeman}, {Friberg}, {Fromm}, {G{\'o}mez}, {Galison}, {Gammie}, {Garc{\'\i}a}, {Gentaz}, {Georgiev}, {Goddi}, {Gold}, {Gu}, {Gurwell}, {Hada}, {Hecht}, {Hesper}, {Ho}, {Ho}, {Honma}, {Huang}, {Huang}, {Hughes}, {Ikeda}, {Inoue}, {Issaoun}, {James}, {Jannuzi}, {Janssen}, {Jeter}, {Jiang}, {Johnson}, {Jorstad}, {Jung}, {Karami}, {Karuppusamy}, {Kawashima}, {Keating}, {Kettenis}, {Kim}, {Kim}, {Kim},
  {Kino}, {Koay}, {Koch}, {Koyama}, {Kramer}, {Kramer}, {Krichbaum}, {Kuo}, {Lauer}, {Lee}, {Li}, {Li}, {Lindqvist}, {Liu}, {Liuzzo}, {Lo}, {Lobanov}, {Loinard}, {Lonsdale}, {Lu}, {MacDonald}, {Mao}, {Markoff}, {Marrone}, {Marscher}, {Mart{\'\i}-Vidal}, {Matsushita}, {Matthews}, {Medeiros}, {Menten}, {Mizuno}, {Mizuno}, {Moran}, {Moriyama}, {Moscibrodzka}, {M{\"u}ller}, {Nagai}, {Nagar}, {Nakamura}, {Narayan}, {Narayanan}, {Natarajan}, {Neri}, {Ni}, {Noutsos}, {Okino}, {Olivares}, {Oyama}, {{\"O}zel}, {Palumbo}, {Patel}, {Pen}, {Pesce}, {Pi{\'e}tu}, {Plambeck}, {PopStefanija}, {Porth}, {Prather}, {Preciado-L{\'o}pez}, {Psaltis}, {Pu}, {Ramakrishnan}, {Rao}, {Rawlings}, {Raymond}, {Rezzolla}, {Ripperda}, {Roelofs}, {Rogers}, {Ros}, {Rose}, {Roshanineshat}, {Rottmann}, {Roy}, {Ruszczyk}, {Ryan}, {Rygl}, {S{\'a}nchez}, {S{\'a}nchez-Arguelles}, {Sasada}, {Savolainen}, {Schloerb}, {Schuster}, {Shao}, {Shen}, {Small}, {Sohn}, {SooHoo}, {Tazaki}, {Tiede}, {Tilanus}, {Titus}, {Toma}, {Torne}, {Trent}, {Trippe},
  {Tsuda}, {van Bemmel}, {van Langevelde}, {van Rossum}, {Wagner}, {Wardle}, {Weintroub}, {Wex}, {Wharton}, {Wielgus}, {Wong}, {Wu}, {Young}, {Young}, \& {Younsi}}]{2019ApJ...875L...4E}
{Event Horizon Telescope Collaboration}, {Akiyama}, K., {Alberdi}, A., {et~al.} 2019, \bibinfo{title}{{First M87 Event Horizon Telescope Results. IV. Imaging the Central Supermassive Black Hole},} \apjl, 875, L4, \dodoi{10.3847/2041-8213/ab0e85}

\bibitem[{D. {Foreman-Mackey} {et~al.}(2013){Foreman-Mackey}, {Hogg}, {Lang}, \& {Goodman}}]{Foreman2013}
{Foreman-Mackey}, D., {Hogg}, D.~W., {Lang}, D., \& {Goodman}, J. 2013, \bibinfo{title}{{emcee: The MCMC Hammer},} \pasp, 125, 306, \dodoi{10.1086/670067}

\bibitem[{G. {Foster}(1996){Foster}}]{1996AJ....112.1709F}
{Foster}, G. 1996, \bibinfo{title}{{Wavelets for period analysis of unevenly sampled time series},} \aj, 112, 1709, \dodoi{10.1086/118137}

\bibitem[{M.~J. {Graham} {et~al.}(2017){Graham}, {Djorgovski}, {Drake}, {Stern}, {Mahabal}, {Glikman}, {Larson}, \& {Christensen}}]{2017MNRAS.470.4112G}
{Graham}, M.~J., {Djorgovski}, S.~G., {Drake}, A.~J., {et~al.} 2017, \bibinfo{title}{{Understanding extreme quasar optical variability with CRTS - I. Major AGN flares},} \mnras, 470, 4112, \dodoi{10.1093/mnras/stx1456}

\bibitem[{A.~D. {Hincks} {et~al.}(2025){Hincks}, {Ma}, {Naess}, {Kiehlmann}, {Mr{\'o}z}, {Bond}, {Devlin}, {Dunkley}, {Foster}, {Graham}, {Guan}, {Herv{\'\i}as-Caimapo}, {Hood}, {Niemack}, {Orlowski-Scherer}, {Page}, {Partridge}, {Readhead}, {Sif{\'o}n}, {Staggs}, \& {Vargas}}]{2025arXiv250404278H}
{Hincks}, A.~D., {Ma}, X., {Naess}, S.~K., {et~al.} 2025, \bibinfo{title}{{Atacama Cosmology Telescope: Observations of supermassive black hole binary candidates. Strong sinusoidal variations at 95, 147 and 225 GHz in PKS 2131$-$021 and PKS J0805$-$0111},} arXiv e-prints, arXiv:2504.04278, \dodoi{10.48550/arXiv.2504.04278}

\bibitem[{B.~C. {Kelly} {et~al.}(2014){Kelly}, {Becker}, {Sobolewska}, {Siemiginowska}, \& {Uttley}}]{2014ApJ...788...33K}
{Kelly}, B.~C., {Becker}, A.~C., {Sobolewska}, M., {Siemiginowska}, A., \& {Uttley}, P. 2014, \bibinfo{title}{{Flexible and Scalable Methods for Quantifying Stochastic Variability in the Era of Massive Time-domain Astronomical Data Sets},} \apj, 788, 33, \dodoi{10.1088/0004-637X/788/1/33}

\bibitem[{S. {Kiehlmann} {et~al.}(2025){Kiehlmann}, {de la Parra}, {Sullivan}, {Synani}, {Liodakis}, {Mr{\'o}z}, {N{\ae}ss}, {Readhead}, {Begelman}, {Blandford}, {Chatziioannou}, {Ding}, {Graham}, {Harrison}, {Homan}, {Hovatta}, {Kulkarni}, {Lister}, {Maiolino}, {Max-Moerbeck}, {Molina}, {O'Dea}, {Pavlidou}, {Pearson}, {Aller}, {Lawrence}, {Lazio}, {O'Neill}, {Prince}, {Ravi}, {Reeves}, {Tassis}, {Vallisneri}, \& {Zensus}}]{2025ApJ...985...59K}
{Kiehlmann}, S., {de la Parra}, P.~V., {Sullivan}, A.~G., {et~al.} 2025, \bibinfo{title}{{PKS 2131‑021{\textemdash}Discovery of Strong Coherent Sinusoidal Variations from Radio to Optical Frequencies: Compelling Evidence for a Blazar Supermassive Black Hole Binary},} \apj, 985, 59, \dodoi{10.3847/1538-4357/adc567}

\bibitem[{J.~H. {Krolik} {et~al.}(2019){Krolik}, {Volonteri}, {Dubois}, \& {Devriendt}}]{Krolik2019ApJ}
{Krolik}, J.~H., {Volonteri}, M., {Dubois}, Y., \& {Devriendt}, J. 2019, \bibinfo{title}{{Population Estimates for Electromagnetically Distinguishable Supermassive Binary Black Holes},} \apj, 879, 110, \dodoi{10.3847/1538-4357/ab24c9}

\bibitem[{D. {Lai} \& D.~J. {Mu{\~n}oz}(2023){Lai} \& {Mu{\~n}oz}}]{2023ARA&A..61..517L}
{Lai}, D., \& {Mu{\~n}oz}, D.~J. 2023, \bibinfo{title}{{Circumbinary Accretion: From Binary Stars to Massive Binary Black Holes},} \araa, 61, 517, \dodoi{10.1146/annurev-astro-052622-022933}

\bibitem[{X.-P. {Li} {et~al.}(2023){Li}, {Yang}, {Cai}, {Song}, {Yang}, \& {Shan}}]{2023RAA....23i5010L}
{Li}, X.-P., {Yang}, H.-Y., {Cai}, Y., {et~al.} 2023, \bibinfo{title}{{A Quasi-periodic Oscillation of 4.6 yr in the Radio Light Curves of Blazar PKS 0607-157},} Research in Astronomy and Astrophysics, 23, 095010, \dodoi{10.1088/1674-4527/ace091}

\bibitem[{G. {Liao} {et~al.}(2025){Liao}, {Chen}, {Zheng}, {Liu}, \& {Zhang}}]{2025A&A...698A.265L}
{Liao}, G., {Chen}, X., {Zheng}, Q., {Liu}, Y., \& {Zhang}, X. 2025, \bibinfo{title}{{Optical quasi periodic oscillations with dual periodicities of 1103 days and 243 days in the blue quasar SDSS J100438.8+151056},} \aap, 698, A265, \dodoi{10.1051/0004-6361/202553747}

\bibitem[{M.~L. {Lister} {et~al.}(2021){Lister}, {Homan}, {Kellermann}, {Kovalev}, {Pushkarev}, {Ros}, \& {Savolainen}}]{2021ApJ...923...30L}
{Lister}, M.~L., {Homan}, D.~C., {Kellermann}, K.~I., {et~al.} 2021, \bibinfo{title}{{Monitoring of Jets in Active Galactic Nuclei with VLBA Experiments. XVIII. Kinematics and Inner Jet Evolution of Bright Radio-loud Active Galaxies},} \apj, 923, 30, \dodoi{10.3847/1538-4357/ac230f}

\bibitem[{F.~K. {Liu} {et~al.}(2006){Liu}, {Zhao}, \& {Wu}}]{2006ApJ...650..749L}
{Liu}, F.~K., {Zhao}, G., \& {Wu}, X.-B. 2006, \bibinfo{title}{{Harmonic QPOs and Thick Accretion Disk Oscillations in the BL Lacertae Object AO 0235+164},} \apj, 650, 749, \dodoi{10.1086/507267}

\bibitem[{N.~R. {Lomb}(1976){Lomb}}]{1976Ap&SS..39..447L}
{Lomb}, N.~R. 1976, \bibinfo{title}{{Least-Squares Frequency Analysis of Unequally Spaced Data},} \apss, 39, 447, \dodoi{10.1007/BF00648343}

\bibitem[{A. {Mainzer} {et~al.}(2011){Mainzer}, {Bauer}, {Grav}, {Masiero}, {Cutri}, {Dailey}, {Eisenhardt}, {McMillan}, {Wright}, {Walker}, {Jedicke}, {Spahr}, {Tholen}, {Alles}, {Beck}, {Brandenburg}, {Conrow}, {Evans}, {Fowler}, {Jarrett}, {Marsh}, {Masci}, {McCallon}, {Wheelock}, {Wittman}, {Wyatt}, {DeBaun}, {Elliott}, {Elsbury}, {Gautier}, {Gomillion}, {Leisawitz}, {Maleszewski}, {Micheli}, \& {Wilkins}}]{2011ApJ...731...53M}
{Mainzer}, A., {Bauer}, J., {Grav}, T., {et~al.} 2011, \bibinfo{title}{{Preliminary Results from NEOWISE: An Enhancement to the Wide-field Infrared Survey Explorer for Solar System Science},} \apj, 731, 53, \dodoi{10.1088/0004-637X/731/1/53}

\bibitem[{L. {Mao} \& X. {Zhang}(2024){Mao} \& {Zhang}}]{2024MNRAS.531.3927M}
{Mao}, L., \& {Zhang}, X. 2024, \bibinfo{title}{{A radio quasi-periodic oscillation in the blazar PKS J2156-0037},} \mnras, 531, 3927, \dodoi{10.1093/mnras/stae1380}

\bibitem[{M.~T. {Miles} {et~al.}(2023){Miles}, {Shannon}, {Bailes}, {Reardon}, {Keith}, {Cameron}, {Parthasarathy}, {Shamohammadi}, {Spiewak}, {van Straten}, {Buchner}, {Camilo}, {Geyer}, {Karastergiou}, {Kramer}, {Serylak}, {Theureau}, \& {Venkatraman Krishnan}}]{2023MNRAS.519.3976M}
{Miles}, M.~T., {Shannon}, R.~M., {Bailes}, M., {et~al.} 2023, \bibinfo{title}{{The MeerKAT Pulsar Timing Array: first data release},} \mnras, 519, 3976, \dodoi{10.1093/mnras/stac3644}

\bibitem[{S. {O'Neill} {et~al.}(2022){O'Neill}, {Kiehlmann}, {Readhead}, {Aller}, {Blandford}, {Liodakis}, {Lister}, {Mr{\'o}z}, {O'Dea}, {Pearson}, {Ravi}, {Vallisneri}, {Cleary}, {Graham}, {Grainge}, {Hodges}, {Hovatta}, {L{\"a}hteenm{\"a}ki}, {Lamb}, {Lazio}, {Max-Moerbeck}, {Pavlidou}, {Prince}, {Reeves}, {Tornikoski}, {Vergara de la Parra}, \& {Zensus}}]{2022ApJ...926L..35O}
{O'Neill}, S., {Kiehlmann}, S., {Readhead}, A.~C.~S., {et~al.} 2022, \bibinfo{title}{{The Unanticipated Phenomenology of the Blazar PKS 2131-021: A Unique Supermassive Black Hole Binary Candidate},} \apjl, 926, L35, \dodoi{10.3847/2041-8213/ac504b}

\bibitem[{T.~J. {Pearson} \& A.~C.~S. {Readhead}(1984){Pearson} \& {Readhead}}]{1984ARA&A..22...97P}
{Pearson}, T.~J., \& {Readhead}, A.~C.~S. 1984, \bibinfo{title}{{Image Formation by Self-Calibration in Radio Astronomy},} \araa, 22, 97, \dodoi{10.1146/annurev.aa.22.090184.000525}

\bibitem[{A.~B. {Pushkarev} {et~al.}(2023){Pushkarev}, {Aller}, {Aller}, {Homan}, {Kovalev}, {Lister}, {Pashchenko}, {Savolainen}, \& {Zobnina}}]{2023MNRAS.520.6053P}
{Pushkarev}, A.~B., {Aller}, H.~D., {Aller}, M.~F., {et~al.} 2023, \bibinfo{title}{{MOJAVE - XX. Persistent linear polarization structure in parsec-scale AGN jets},} \mnras, 520, 6053, \dodoi{10.1093/mnras/stad525}

\bibitem[{A.~C.~S. {Readhead} {et~al.}(1989){Readhead}, {Lawrence}, {Myers}, {Sargent}, {Hardebeck}, \& {Moffet}}]{1989ApJ...346..566R}
{Readhead}, A.~C.~S., {Lawrence}, C.~R., {Myers}, S.~T., {et~al.} 1989, \bibinfo{title}{{A Limit on the Anisotropy of the Microwave Background Radiation on Arcminute Scales},} \apj, 346, 566, \dodoi{10.1086/168039}

\bibitem[{G.-W. {Ren} {et~al.}(2021{\natexlab{a}}){Ren}, {Ding}, {Zhang}, {Xue}, {Zhang}, {Xiong}, {Li}, \& {Li}}]{2021MNRAS.506.3791R}
{Ren}, G.-W., {Ding}, N., {Zhang}, X., {et~al.} 2021{\natexlab{a}}, \bibinfo{title}{{Detection of a possible high-confidence radio quasi-periodic oscillation in the BL Lac PKS J2134-0153},} \mnras, 506, 3791, \dodoi{10.1093/mnras/stab1739}

\bibitem[{G.-W. {Ren} {et~al.}(2021{\natexlab{b}}){Ren}, {Zhang}, {Zhang}, {Ding}, {Yang}, {Li}, {Yan}, \& {Xu}}]{2021RAA....21...75R}
{Ren}, G.-W., {Zhang}, H.-J., {Zhang}, X., {et~al.} 2021{\natexlab{b}}, \bibinfo{title}{{Detection of a high-confidence quasi-periodic oscillation in radio light curve of the high redshift FSRQ PKS J0805-0111},} Research in Astronomy and Astrophysics, 21, 075, \dodoi{10.1088/1674-4527/21/3/075}

\bibitem[{G.~T. {Richards} {et~al.}(2009){Richards}, {Myers}, {Gray}, {Riegel}, {Nichol}, {Brunner}, {Szalay}, {Schneider}, \& {Anderson}}]{2009ApJS..180...67R}
{Richards}, G.~T., {Myers}, A.~D., {Gray}, A.~G., {et~al.} 2009, \bibinfo{title}{{Efficient Photometric Selection of Quasars from the Sloan Digital Sky Survey. II. \raisebox{-0.5ex}\textasciitilde1,000,000 Quasars from Data Release 6},} \apjs, 180, 67, \dodoi{10.1088/0067-0049/180/1/67}

\bibitem[{J.~L. {Richards} {et~al.}(2011){Richards}, {Max-Moerbeck}, {Pavlidou}, {King}, {Pearson}, {Readhead}, {Reeves}, {Shepherd}, {Stevenson}, {Weintraub}, {Fuhrmann}, {Angelakis}, {Zensus}, {Healey}, {Romani}, {Shaw}, {Grainge}, {Birkinshaw}, {Lancaster}, {Worrall}, {Taylor}, {Cotter}, \& {Bustos}}]{2011ApJS..194...29R}
{Richards}, J.~L., {Max-Moerbeck}, W., {Pavlidou}, V., {et~al.} 2011, \bibinfo{title}{{Blazars in the Fermi Era: The OVRO 40 m Telescope Monitoring Program},} \apjs, 194, 29, \dodoi{10.1088/0067-0049/194/2/29}

\bibitem[{M. {Roy} {et~al.}(2000){Roy}, {Papadakis}, {Ramos-Col{\'o}n}, {Sambruna}, {Tsinganos}, {Papamastorakis}, \& {Kafatos}}]{2000ApJ...545..758R}
{Roy}, M., {Papadakis}, I.~E., {Ramos-Col{\'o}n}, E., {et~al.} 2000, \bibinfo{title}{{The Recent High State of the BL Lacertae Object AO 0235 and Cross-Correlations between Optical and Radio Bands},} \apj, 545, 758, \dodoi{10.1086/317842}

\bibitem[{J.~D. {Scargle}(1982){Scargle}}]{1982ApJ...263..835S}
{Scargle}, J.~D. 1982, \bibinfo{title}{{Studies in astronomical time series analysis. II. Statistical aspects of spectral analysis of unevenly spaced data.},} \apj, 263, 835, \dodoi{10.1086/160554}

\bibitem[{M. {Seilmayer} {et~al.}(2020){Seilmayer}, {Garcia Gonzalez}, \& {Wondrak}}]{2020arXiv200110200S}
{Seilmayer}, M., {Garcia Gonzalez}, F., \& {Wondrak}, T. 2020, \bibinfo{title}{{The Multivariate Extension of the Lomb-Scargle Method},} arXiv e-prints, arXiv:2001.10200, \dodoi{10.48550/arXiv.2001.10200}

\bibitem[{A. {Sharma} {et~al.}(2023){Sharma}, {Prince}, \& {Bose}}]{2023arXiv231212623S}
{Sharma}, A., {Prince}, R., \& {Bose}, D. 2023, \bibinfo{title}{{Detection of gamma-ray quasi-periodic oscillations in non-blazar AGN PKS 0521-36},} arXiv e-prints, arXiv:2312.12623, \dodoi{10.48550/arXiv.2312.12623}

\bibitem[{M.~C. {Shepherd}(1997){Shepherd}}]{1997ASPC..125...77S}
{Shepherd}, M.~C. 1997, \bibinfo{title}{{Difmap: an Interactive Program for Synthesis Imaging},} in Astronomical Society of the Pacific Conference Series, Vol. 125, Astronomical Data Analysis Software and Systems VI, ed. G.~{Hunt} \& H.~{Payne}, 77

\bibitem[{M.~C. {Shepherd} {et~al.}(1994){Shepherd}, {Pearson}, \& {Taylor}}]{1994BAAS...26..987S}
{Shepherd}, M.~C., {Pearson}, T.~J., \& {Taylor}, G.~B. 1994, \bibinfo{title}{{DIFMAP: an interactive program for synthesis imaging.},} in Bulletin of the American Astronomical Society, Vol.~26, 987--989

\bibitem[{E. {Sobacchi} {et~al.}(2017){Sobacchi}, {Sormani}, \& {Stamerra}}]{2017MNRAS.465..161S}
{Sobacchi}, E., {Sormani}, M.~C., \& {Stamerra}, A. 2017, \bibinfo{title}{{A model for periodic blazars},} \mnras, 465, 161, \dodoi{10.1093/mnras/stw2684}

\bibitem[{Y. {Tang} {et~al.}(2018){Tang}, {Haiman}, \& {MacFadyen}}]{Tang2018MNRAS}
{Tang}, Y., {Haiman}, Z., \& {MacFadyen}, A. 2018, \bibinfo{title}{{The late inspiral of supermassive black hole binaries with circumbinary gas discs in the LISA band},} \mnras, 476, 2249, \dodoi{10.1093/mnras/sty423}

\bibitem[{A. {Tripathi} {et~al.}(2021){Tripathi}, {Gupta}, {Aller}, {Wiita}, {Bambi}, {Aller}, \& {Gu}}]{2021MNRAS.501.5997T}
{Tripathi}, A., {Gupta}, A.~C., {Aller}, M.~F., {et~al.} 2021, \bibinfo{title}{{Quasi-periodic oscillations in the long-term radio light curves of the blazar AO 0235+164},} \mnras, 501, 5997, \dodoi{10.1093/mnras/stab058}

\bibitem[{J.~T. {VanderPlas} \& {\v{Z}}. {Ivezi{\'c}}(2015){VanderPlas} \& {Ivezi{\'c}}}]{2015ApJ...812...18V}
{VanderPlas}, J.~T., \& {Ivezi{\'c}}, {\v{Z}}. 2015, \bibinfo{title}{{Periodograms for Multiband Astronomical Time Series},} \apj, 812, 18, \dodoi{10.1088/0004-637X/812/1/18}

\bibitem[{J.-Y. {Wang} {et~al.}(2014){Wang}, {An}, {Baan}, \& {Lu}}]{2014MNRAS.443...58W}
{Wang}, J.-Y., {An}, T., {Baan}, W.~A., \& {Lu}, X.-L. 2014, \bibinfo{title}{{Periodic radio variabilities of the blazar 1156+295: harmonic oscillations},} \mnras, 443, 58, \dodoi{10.1093/mnras/stu1135}

\bibitem[{J.-y. {Wang} {et~al.}(2017){Wang}, {Lu}, {Zhao}, {Dong}, {Lao}, {Lu}, {Wei}, {Wu}, \& {An}}]{2017ChA&A..41...42W}
{Wang}, J.-y., {Lu}, X.-l., {Zhao}, Q.-w., {et~al.} 2017, \bibinfo{title}{{Periodicity Analysis of X-ray Light Curves of SS 433},} \caa, 41, 42, \dodoi{10.1016/j.chinastron.2017.01.005}

\bibitem[{Z.~R. {Weaver} {et~al.}(2022){Weaver}, {Jorstad}, {Marscher}, {Morozova}, {Troitsky}, {Agudo}, {G{\'o}mez}, {L{\"a}hteenm{\"a}ki}, {Tammi}, \& {Tornikoski}}]{2022ApJS..260...12W}
{Weaver}, Z.~R., {Jorstad}, S.~G., {Marscher}, A.~P., {et~al.} 2022, \bibinfo{title}{{Kinematics of Parsec-scale Jets of Gamma-Ray Blazars at 43 GHz during 10 yr of the VLBA-BU-BLAZAR Program},} \apjs, 260, 12, \dodoi{10.3847/1538-4365/ac589c}

\bibitem[{E.~L. {Wright} {et~al.}(2019){Wright}, {Eisenhardt}, {Mainzer}, {Ressler}, {Cutri}, {Jarrett}, {Kirkpatrick}, {Padgett}, {McMillan}, {Skrutskie}, {Stanford}, {Cohen}, {Walker}, {Mather}, {Leisawitz}, {Gautier}, {McLean}, {Benford}, {Lonsdale}, {Blain}, {Mendez}, {Irace}, {Duval}, {Liu}, {Royer}, {Heinrichsen}, {Howard}, {Shannon}, {Kendall}, {Walsh}, {Larsen}, {Cardon}, {Schick}, {Schwalm}, {Abid}, {Fabinsky}, {Naes}, \& {Tsai}}]{2019ipac.data...I1W}
{Wright}, E.~L., {Eisenhardt}, P. R.~M., {Mainzer}, A.~K., {et~al.} 2019, {AllWISE Source Catalog},, NASA IPAC DataSet, IRSA1 \dodoi{10.26131/IRSA1}

\bibitem[{M. {Zechmeister} \& M. {K{\"u}rster}(2009){Zechmeister} \& {K{\"u}rster}}]{2009AandA...496..577Z}
{Zechmeister}, M., \& {K{\"u}rster}, M. 2009, \bibinfo{title}{{The generalised Lomb-Scargle periodogram. A new formalism for the floating-mean and Keplerian periodograms},} \aap, 496, 577, \dodoi{10.1051/0004-6361:200811296}

\end{thebibliography}
\bibliographystyle{aasjournalv7}





\end{document}